\documentclass[onecolumn,showpacs,preprintnumbers,amsmath,amssymb]{revtex4-2}
\usepackage{amsmath}
\usepackage{amssymb}
\usepackage{graphicx}
\usepackage{bm}
\usepackage{epstopdf}
\usepackage{slashed}
\usepackage{color}
\usepackage{float} 
\usepackage{caption}
\usepackage{subcaption}
\usepackage[normalem]{ulem}
\begin{document}

\title{Point-Like Charges and Dirichlet Boundary Conditions in a Nonlocal Scalar Field Theory}

\author{L.H.C. Borges}
\email{luizhenrique.borges@ufla.br}

\author{Cleverson Filgueiras}
\email{cleverson.filgueiras@ufla.br}

\affiliation{Departamento de Física, Universidade Federal de Lavras, Caixa Postal 3037, 37200-900 Lavras-MG, Brazil.}

\author{F.A. Barone}
\email{fbarone@unifei.edu.br}

\affiliation{Instituto de Física e Química,
Universidade Federal de Itajub\'a, Av. BPS 1303 Itajubá-MG, Brazil.}

\author{A. A. Nogueira}
\email{andsogueira@hotmail.com}

\affiliation{Departamento de Física, Universidade do Estado de Santa Catarina, Joinville, SC, 89219-710, Brasil.}

\begin{abstract}
In this paper, we investigate novel classical non-local effects in scalar field theory arising from the presence of external sources and a material boundary. Specifically, we consider a model in which the standard Klein–Gordon field theory is modified by a non-local term. First, we analyze the interaction between stationary field sources, modeled as point-like scalar charges. Then, we examine the model in the presence of a Dirichlet plane, computing both the scalar field propagator and the interaction force between the Dirichlet plane and a point-like scalar charge. Additionally, we demonstrate the failure of the image method in this non-local context and compare our results with those obtained in non-local electrodynamics and in the standard Klein–Gordon field theory.
\end{abstract}
\maketitle
\section{Introduction}
\label{I}

Non-local field theories have been introduced to improve the ultraviolet behavior of quantized theories and to address the ghost problem. Furthermore, such theories also provide promising frameworks for exploring alternative approaches to unresolved problems in high-energy physics. Consequently, non-local theories have received considerable attention in the literature. For instance, studies have examined the role of non-locality in quantum theories of gravity~\cite{Q1,Q11,Q2,Q3,Q4,G1,G2}, as well as in the search for non-local effects within both quantum~\cite{QU1,QU2,QU3,QU4} and classical electrodynamics~\cite{CL1,CL2,CL3,CL4}. Additionally, attention has been given to the implications of non-locality in scenarios involving Lorentz violation~\cite{LV1,LV2}, condensed matter systems~\cite{C1,C2,C3}, kink solutions~\cite{Kink}, the Higgs mechanism~\cite{Hig1,Hig2}, supersymmetry~\cite{SS1,SS2}, and in the context of spin-$1/2$ theory~\cite{SP12}, and the Casimir effect \cite{CasimirPRA2025,CasimirCQG2019,CasimirAIPconf2005}, among various other areas of interest.

It is well established that field theories incorporating boundary conditions have a wide range of applications across various branches of physics. In particular, phenomena associated with the presence of material boundaries in field theories have attracted considerable attention~\cite{Casimir,Mostepanenko,BordagLivro,Milton,MiltonLivro,CasimirKR,Ravndal,FTAB,Scharnhorst,ScharnhorstBarton,ScharnhorstNos,VacuoQED,Emissaoexp,MagneticoVacuoEscalar,MagneticoVacuoQED,AtomoParede,AtomoCunha,LHCBFABplate,BC14,BC15,BC16}, to cite just a few examples. An alternative approach to addressing such problems involves the study of models where fields are coupled to external potentials defined on surfaces~\cite{Bord,Jaff,Milton,GTFABFEB,FABFEB1,FABFEB2,LHCBAFFFAB,BRW,BC8,BC9}. Moreover, the investigation of interactions between external sources has unveiled several interesting features across different theoretical frameworks~\cite{LHCBAFFFAB,F1,F2,F3,F4,F5,F6,F7,F8,F9,F10}.

Within this broad context, it is useful to distinguish between different classes of nonlocal operators, which lead to qualitatively distinct physical effects. In particular, exponential operators of the form $e^{\Box/M^2}$, which naturally arise in string field theory and in ultraviolet-complete extensions of gravity, are known to provide a soft ultraviolet behavior~\cite{Calcagni2007,Calcagni2008,Koshelev2016}. In contrast, rational nonlocal operators of the form $(\Box + m^2)^{-1}$, closely related to Podolsky-like higher-derivative theories, primarily affect the infrared sector of the theory. Such structures are capable of generating nontrivial long-range modifications, including linear contributions to the interaction energy analogous to the Cornell confining potential~\cite{Eichten1978,Gaete1999,Gaete2001,GaeteSpallucci2008}.

Furthermore, the interplay between nonlocality and boundary conditions has been investigated in the context of Casimir-type systems, where additional modifications arise due to the presence of surfaces and external constraints~\cite{Fosco2011}. These features highlight that different realizations of nonlocality may control distinct physical regimes, making rational nonlocal models particularly suitable for exploring infrared effects in the presence of external sources and boundaries.

Regarding the presence of external sources and boundary conditions in non-local electrodynamics, Ref.~\cite{CL4} considered a non-local model (proposed in Refs.~\cite{COR1,COR2,COR3}) that describes the phenomenological Cornell confining potential~\cite{CORNELL}. In this context, several non-local classical effects were obtained, arising from the interaction between external sources and from the presence of a perfectly conducting surface. However, to the best of the authors' knowledge, non-local scalar field theories have not yet been considered in the presence of external sources and boundary conditions.

A natural question that arises in this context concerns the modifications introduced by non-locality in the interaction between point-like sources, as well as the changes experienced by the non-local scalar field propagator in the presence of a single Dirichlet plane, and the influence of such a surface on the dynamics of point-like field sources. Moreover, studies of this kind have been scarcely explored in the context of non-local theories.

These issues are important for several reasons. The scalar field is a versatile tool for modeling a wide range of condensed matter systems. External sources and potentials coupled to it are used to simulate the presence of defects and/or boundaries in material samples. 
Moreover, scalar field systems are often used in toy models to mimic the behavior of vector fields in certain scenarios.

This paper is devoted to this subject, in which we investigate certain aspects of a non-local scalar field theory in the presence of point-like charges and near a single Dirichlet plane. We consider a model where a non-local term is added to the standard Klein--Gordon Lagrangian. This term corresponds to the scalar field version of the one introduced in Refs.~\cite{CL4,COR1,COR2,COR3} for non-local electrodynamics.

The model considered in this paper can be regarded as an effective theory obtained by integrating out a massive scalar field linearly coupled to a second, lighter scalar field. Models with coupled scalar fields are of particular interest in condensed matter systems. In this context, one may mention, for instance, phonon dynamics in binary superfluids \cite{PRD125024}, as well as the dynamics of coupled phonons in solids \cite{PRD105009}, the modeling of phase transitions \cite{PRD036019}, effective descriptions of dissipative effects \cite{SCIREP2019}, and symmetry breaking and restoration under external fields \cite{PRD096011}.

Despite this, there is still a gap in the literature regarding effects associated with the presence of impurities and boundaries in material media. The results presented in this work may serve as a starting point for addressing this issue, particularly in the regime where one field is lighter than the other (the soft-phonon regime).

More specifically, in Sect.~\ref{II} we investigate the interaction between stationary point-like scalar charges for various configurations. In Sect.~\ref{III}, we compute the propagator of the non-local scalar field in the presence of a Dirichlet plane under different scenarios. In Sect.~\ref{IV}, using the previously obtained propagators, we calculate both the interaction energy and the interaction force between a point-like scalar charge and the Dirichlet plane. Exact analytical results are obtained for certain special cases, while a numerical analysis is performed for more general configurations. We compare the interaction forces with those obtained in the free theory (i.e., without the Dirichlet plane), and verify that the image method is not applicable in our model with Dirichlet boundary conditions. Additionally, we compare the results presented throughout the paper with those found in the context of non-local electrodynamics and the standard Klein--Gordon field theory. Finally, Sect.~\ref{V} presents our concluding remarks and final considerations.

Throughout the paper, we work in a (3 + 1) Minkowski space-time with metric $\eta^{\mu\nu}=\left(1,-1,-1,-1\right)$.

\section{Point-like scalar charges}
\label{II}

In this section, we investigate nonlocal effects in a scalar field theory induced by the presence of stationary point-like scalar sources. In particular, we focus on how the nonlocal structure of the model modifies the interaction between sources and the corresponding propagator behavior. The model is described by the following Lagrangian density:
\begin{eqnarray}
 \label{model1}
 {\cal{L}}=\frac{1}{2} \partial_{\mu}\phi\,\partial^{\mu}\phi
 -\frac{m^{2}}{2}\,\partial_{\mu}\phi\,\frac{1}{\Box+m^{2}}\,\partial^{\mu}\phi
 -\frac{1}{2}M^{2}\phi^{2}
 +J\phi \ ,
\end{eqnarray}
where $\phi$ is a real scalar field, $M$ denotes its mass, $J$ is an external source describing the point-like charges, and $m$ is a parameter with dimensions of mass controlling the nonlocal sector of the theory.

To better understand the role played by the nonlocal sector, it is useful to compare the present model with the exponential nonlocal operators often considered in the literature. A typical example is given by operators of the form
\begin{equation}
\label{eqexpo}
{\cal{L}}=\frac{1}{2}\partial_{\mu}\phi e^{-\,\frac{m^2}{\Box+m^2}}\partial^{\mu}\phi
 -\frac{1}{2}M^{2}\phi^{2}
 +J\phi 
\end{equation}
Assuming that the dimensionless rational operator
\begin{equation}
\frac{m^2}{\Box+m^2}
\end{equation}
is sufficiently small in the regime of interest, one may formally expand the exponential as
\begin{equation}
e^{-\,\frac{m^2}{\Box+m^2}}
=
1
-\frac{m^2}{\Box+m^2}
+\frac{1}{2!}\left(\frac{m^2}{\Box+m^2}\right)^2
-\frac{1}{3!}\left(\frac{m^2}{\Box+m^2}\right)^3
+\cdots .
\end{equation}
This expression shows that an exponential nonlocal structure generates an infinite tower of nonlocal operators built from the kernel $(\Box+m^2)^{-1}$. In the ultraviolet regime, where $|\Box| \gg m^2$, one has
\begin{equation}
\frac{m^2}{\Box+m^2}\approx \frac{m^2}{\Box},
\end{equation}
so that
\begin{equation}
e^{-\,\frac{m^2}{\Box+m^2}}
\approx
e^{-\,\frac{m^2}{\Box}}
\approx
1-\frac{m^2}{\Box}+\cdots .
\end{equation}
Hence, in this limit the nonlocal contribution becomes suppressed and the theory tends to recover a quasi-local behavior. This is in contrast with the usual exponential operators $e^{\Box/M^2}$, which strongly affect the ultraviolet sector. In the present case, the rational operator $(\Box+m^2)^{-1}$ acts mainly in the infrared regime, making the model particularly suitable for investigating long-range modifications of the interaction energy. It is important to note that, to the best of our knowledge, the model introduced in Eq. (\ref{eqexpo}) is novel and has not been discussed in the literature before.

The Lagrangian (\ref{model1}) can be rewritten as
\begin{eqnarray}
\label{model2}   
{\cal{L}}\rightarrow -\frac{1}{2}\phi{\cal{O}}\phi +J \phi \ ,
\end{eqnarray}
with the differential operator defined by
\begin{eqnarray}
\label{opeDif}   
{\cal{O}}=\Box-m^{2}\frac{\Box}{\Box+m^{2}}+M^{2} \ .
\end{eqnarray}
We notice that the propagator $D\left(x,y\right)$ satisfies
\begin{eqnarray}
\label{prop1}    
{\cal{O}}D\left(x,y\right)=\delta^4\left(x-y\right) \ ,
\end{eqnarray}
where we arrive at
\begin{eqnarray}
\label{prop2}   
D\left(x,y\right)=-\int\frac{d^{4}p}{\left(2\pi\right)^{4}}\frac{p^{2}-m^{2}}{p^{4}-M^{2}p^{2}+M^{2}m^{2}} \ e^{-ip\cdot\left(x-y\right)} \ .
\end{eqnarray}
From Eq. (\ref{prop2}), one can show that the model (\ref{model1}) exhibits two massive poles for the squared momentum, given by
\begin{eqnarray}
\label{poleS}   
m_{+}^{2}&=&\frac{M^{2}}{2}\left[1+\left(1-\frac{4m^{2}}{M^{2}}\right)^{1/2}\right] \ , \\
\label{poless}
m_{-}^{2}&=&\frac{M^{2}}{2}\left[1-\left(1-\frac{4m^{2}}{M^{2}}\right)^{1/2}\right] \ .
\end{eqnarray}

In order to avoid tachyonic modes, and vacuum instability, one must impose the condition
\begin{eqnarray}
\label{cond1}   
0\leq\frac{4m^{2}}{M^{2}}\leq 1 \ .
\end{eqnarray}

Since the theory is quadratic in the field variables $\phi$, it can be shown that the contribution of the source $J\left(x\right)$ to the vacuum energy of the system reads \cite{LHCBAFFFAB,F4,F7}
\begin{eqnarray}
\label{E1}   
E=-\frac{1}{2T}\int\int d^{4}x \ d^{4}y \ J\left(x\right)  D\left(x,y\right)J\left(y\right) \ ,
\end{eqnarray}
where $T$ denotes the time variable, and the limit $T \rightarrow \infty$ is understood to be taken at the end of the calculations.

From now on, for the model (\ref{model1}), let us compute the interaction energy between two point-like scalar charges, namely $\sigma_{1}$ and $\sigma_{2}$, fixed at the spatial positions ${\bf{a}}_{1}$ and ${\bf{a}}_{2}$, respectively. Such a system is described by the external source 
\begin{eqnarray}
\label{source1}    
J\left(x\right)=\sigma_{1}\delta^{3}\left({\bf{x}}-{\bf{a}}_{1}\right)+\sigma_{2}\delta^{3}\left({\bf{x}}-{\bf{a}}_{2}\right) \ .
\end{eqnarray}

We first consider the case where $M=0$ (vanishing Klein-Gordon mass), wherein both $m_{+}$ and $m_{-}$ vanish. In this setup, the propagator (\ref{prop2}) reads 
\begin{eqnarray}
\label{prop31}    
D\left(x,y\right)\mid_{M=0}=-\int\frac{d^{4}p}{\left(2\pi\right)^{4}}\frac{p^{2}-m^{2}}{p^{4}}e^{-ip\cdot\left(x-y\right)} \ .
\end{eqnarray}

Substituting (\ref{prop31}) and (\ref{source1}) into (\ref{E1}), discarding the self-interaction contributions (interactions of a given charge with itself), performing the integrals in the following order: $d^{3}{\bf x}$, $d^{3}{\bf y}$, $dx^{0}$, $dp^{0}$, and $dy^{0}$, using the Fourier representation for the Dirac delta function, $\delta(p^{0})=\int dx/(2\pi)\exp(-ipx^{0})$, and identifying the time interval as $T=\int dy^{0}$, we obtain 
\begin{eqnarray}
\label{ECC1}   
E^{CC}\left(M=0\right)=-\sigma_{1}\sigma_{2}\int\frac{d^{3}{\bf p}}{(2\pi)^{3}}\frac{e^{i{\bf p}\cdot{\bf a}}}{{\bf {p}}^{2}}-\sigma_{1}\sigma_{2}m^{2}\int\frac{d^{3}{\bf p}}{(2\pi)^{3}}\frac{e^{i{\bf p}\cdot{\bf a}}}{{\bf {p}}^{4}}\ ,
\end{eqnarray}
which leads to the result
\begin{equation}
\label{ECC2}   
E^{CC}\left(M=0\right)=-\frac{\sigma_{1}\sigma_{2}}{4\pi a}+\frac{\sigma_{1}\sigma_{2}m^{2}}{8\pi}a \ ,
\end{equation}
where we defined ${{\bf {a}}={\bf {a}}_{1}-{\bf {a}}_{2}}$, with $a=\mid{\bf{a}}\mid$, as the distance between the two charges. 
The superscript $CC$ refers to charge–charge interaction.

\begin{figure}[!h]
\centering \includegraphics[scale=0.45]{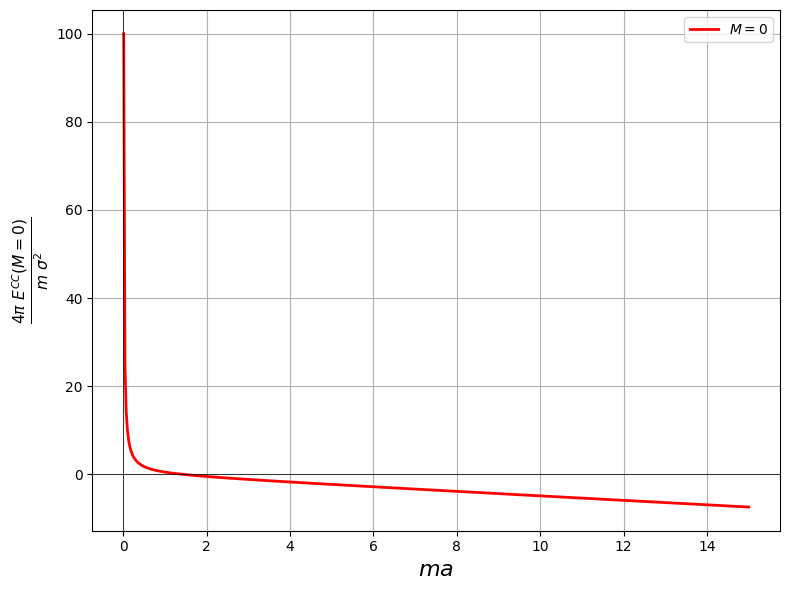} \caption{Energy given by Eq.~(\ref{ECC3}), multiplied by the factor $\frac{4\pi}{\sigma^2 m}$, as a function of the dimensionless parameter $ma$. The energy diverges positively as $ma \to 0$, crosses zero near $ma \approx 2$, and becomes negative for larger values of $ma$.}
\label{grafico1}
\end{figure}

Equation (\ref{ECC2}) gives the interaction energy between two stationary point-like charges for the model (\ref{model1}). The first term on the right-hand side of Eq. (\ref{ECC2}) corresponds to the Coulomb interaction \cite{F7}, while the second one, which depends on the parameter $m$, represents a contribution induced by the non-locality of the model. For the configuration with two scalar charges of opposite signs, $\sigma_{1}=-\sigma_{2}=\sigma$, Eq. (\ref{ECC2}) becomes
\begin{equation}
\label{ECC3}   
E^{CC}\left(M=0\right)=\frac{\sigma^{2}}{4\pi a}\left[1-\frac{1}{2}\left(ma\right)^{2}\right] \ .
\end{equation}
We notice that this result is equivalent to that obtained in Ref. \cite{CL4} for non-local electrodynamics, multiplied by an overall minus sign. Figure (\ref{grafico1}) illustrates the general behavior of this interaction energy.

Figure (\ref{grafico1}) shows that within the range \( 0 < ma < \sqrt{2} \), the interaction energy given by Eq. (\ref{ECC3}) is positive, with its magnitude decreasing as \( ma \) increases. For small values of \( ma \), the energy behaves similarly to a Coulomb potential. At \( ma = \sqrt{2} \), the energy \( E^{CC}\left(M=0\right) \) vanishes, indicating that the \( m \)-dependent contribution in Eq. (\ref{ECC3}) cancels the Coulomb term. Furthermore, for \( ma > \sqrt{2} \), \( E^{CC}\left(M=0\right) \) becomes negative and continues to grow in magnitude as \( ma \) increases. For large \( ma \), the interaction energy (\ref{ECC3}) is predominantly governed by its non-local term, leading to long-range interaction behavior.

\begin{figure}[!h]
\centering \includegraphics[scale=0.45]{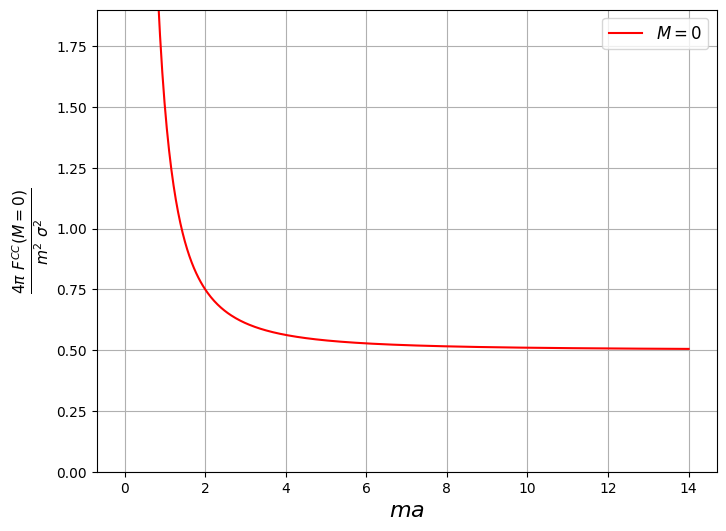} \caption{Interaction force derived from Eq.~(\ref{FCC1}), multiplied by the factor $\frac{4\pi}{\sigma^2 m^2}$, as a function of the dimensionless parameter $ma$. The force diverges positively for $ma \to 0$ and decreases monotonically, approaching a non-zero constant for large $ma$. This behavior reflects the derivative of the energy shown in Fig.~\ref{grafico1}. The force remains positive throughout, indicating a repulsive interaction.}
\label{grafico2}
\end{figure}

The interaction force can be obtained from Eq. (\ref{ECC3}) as follows:
\begin{equation}
\label{FCC1}   
F^{CC}\left(M=0\right)=-\frac{dE^{CC}\left(M=0\right)}{da}=\frac{\sigma^{2}}{4\pi a^{2}}\left[1+\frac{1}{2}\left(ma\right)^{2}\right] \ ,
\end{equation}
where we have the usual Coulomb force modified by a non-local contribution ($m$-dependent contribution). In Fig. (\ref{grafico2}), the general behavior of the force (\ref{FCC1}) is shown as a function of $ma$. 

From Fig. (\ref{grafico2}), we see that the force (\ref{FCC1}) is always repulsive, since the term inside the brackets on the right-hand side is always positive (in Ref. \cite{CL4}, for non-local electrodynamics, a similar expression appears, but with attractive behavior). For small values of $ma$, this force exhibits a Coulomb-like behavior. As $ma$ increases, the interaction force (\ref{FCC1}) approaches a finite value given by
\begin{equation}
\label{FCCLIM}   
\lim_{ma\rightarrow\infty}\frac{[F^{CC}(M=0)]4\pi}{\sigma^{2}m^{2}}=\frac{1}{2} \Rightarrow F^{CC}\left(M=0\right)= \frac{\sigma^{2}}{8\pi}m^{2} \ ,
\end{equation}
which describes a long-range interaction force solely governed by the mass parameter $m$ associated with the non-local contribution.
This behavior differs from that of the usual Klein-Gordon field theory, where the force vanishes as $ma$ increases.

It is also worth mentioning the situation where we have two point-like scalar charges with the same sign, $\sigma_{1}=\sigma_{2}=\sigma$, for which the interaction energy (\ref{ECC3}) reads
\begin{equation}
\label{ECC73}   
E^{CC}\left(M=0\right)=-\frac{\sigma^{2}}{4\pi a}\left[1-\frac{1}{2}\left(ma\right)^{2}\right] \ ,
\end{equation}
which resembles a confinement Cornell-type potential \cite{CORNELL}, commonly used in the study of quark–antiquark interactions. This situation is similar to the one studied in Ref. \cite{CL4}, for non-local electrodynamics.


In the next situation, we consider the setup where $M=2m$, which is the maximum value that the mass of the scalar field can assume without the appearance of tachyonic modes, as can be seen from the conditions (\ref{cond1}). For this configuration, we have two identical massive poles, namely $m_{+}=m_{-}=M/{\sqrt{2}}$. The propagator now reads  
\begin{eqnarray}
\label{prop32}    
D\left(x,y\right)\mid_{M=2m}=-\int\frac{d^{4}p}{\left(2\pi\right)^{4}}\frac{p^{2}-m^{2}}{\left(p^{2}-2m^{2}\right)^{2}} \ e^{-ip\cdot\left(x-y\right)} \ .
\end{eqnarray}

Substituting (\ref{source1}) and (\ref{prop32}) into (\ref{E1}) and performing some manipulations, we obtain
\begin{eqnarray}
\label{ECC4}   
E^{CC}\left(M=2m\right)=-\sigma_{1}\sigma_{2}\left(-{\bf{\nabla}}_{{\bf{a}}}^{2}+m^{2}\right)\int\frac{d^{3}{\bf p}}{(2\pi)^{3}}\frac{e^{i{\bf p}\cdot{\bf a}}}{\left({\bf{p}}^{2}+2m^{2}\right)^{2}} \ ,
\end{eqnarray}
where we defined the differential operator ${\bf{\nabla}}_{{\bf{a}}}=\left(\frac{\partial}{\partial a^{1}}, \frac{\partial}{\partial a^{2}},\frac{\partial}{\partial a^{3}}\right)$.  
Carrying out the calculations, we arrive at
\begin{eqnarray}
\label{ECC5}    
E^{CC}\left(M=2m\right)=-\frac{\sigma_{1}\sigma_{2}}{4\pi a}e^{-\sqrt{2} \ ma}\left(1-\frac{ma}{2{\sqrt{2}}}\right) \ .
\end{eqnarray}

\begin{figure}[!h]
\centering \includegraphics[scale=0.45]{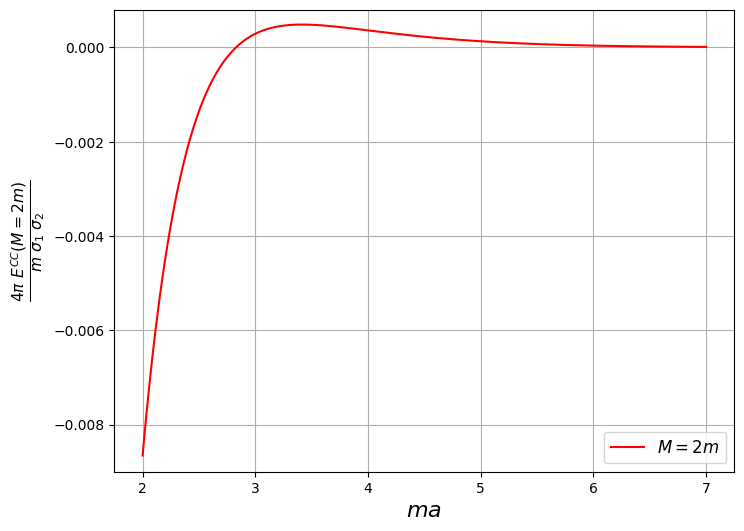} \caption{Interaction energy Eq.~(\ref{ECC5}), multiplied by the factor $\frac{4\pi}{\sigma_1 \sigma_2 m}$, as a function of the dimensionless parameter $ma$, for $M = 2m$. The energy is negative for small $ma$, reaching a minimum near $ma \approx 2$, then increases and crosses zero around $ma \approx 2.8$. For larger $ma$, the energy becomes slightly positive and tends to zero. This behavior indicates a transition from attractive to repulsive interaction.}
\label{grafico3}
\end{figure}

Eq. (\ref{ECC5}) gives the interaction energy between two point-like charges for the case where $M=2m$. The first term in parentheses on the right-hand side corresponds to the standard Yukawa potential (short-range interaction) \cite{F7}. The second term can be interpreted as the correction to the Yukawa interaction induced by the non-local term. In Fig. (\ref{grafico3}), we show a plot of the energy (\ref{ECC5}) multiplied by $\frac{4\pi}{\sigma_{1}\sigma_{2}m}$. 

Figure~(\ref{grafico3}) shows that in the interval \( 0 < ma < 2\sqrt{2} \), the interaction energy~(\ref{ECC5}) is negative and increases with \( ma \). For small values of \( ma \), the energy exhibits Yukawa-like behavior; in the limit \( ma \rightarrow 0 \), the magnitude of the interaction energy~(\ref{ECC5}) diverges. At \( ma = 2\sqrt{2} \), we have \( E^{CC}(M = 2m) = 0 \), indicating that the non-local contribution to the potential~(\ref{ECC5}) becomes equal in magnitude to the Yukawa term.

For \( ma > 2\sqrt{2} \), the energy becomes positive, and the non-local term dominates over the Yukawa one. At large \( ma \), the interaction energy is entirely governed by its non-local component, resulting in a long-range interaction.

Additionally, we find that in the interval \( 0 < ma < 2 + \sqrt{2} \), the energy increases with \( ma \), reaching a maximum at \( ma = 2 + \sqrt{2} \). Beyond this point, the energy decreases as \( ma \) grows, approaching zero in the limit \( ma \rightarrow \infty \). This implies that \( ma = 2 + \sqrt{2} \) corresponds to an unstable equilibrium point of the system. Such behavior has no counterpart in conventional theories without non-locality.

Additionally, the interaction force between the two charges reads
\begin{eqnarray}
\label{FCCS2}    
F^{CC}\left(M=2m\right)=-\frac{\sigma_{1}\sigma_{2}}{4\pi a^{2}}\left(1+{\sqrt{2}} \ ma\right)e^{-\sqrt{2} \ ma}\left[1-\frac{1}{2}\frac{\left(ma\right)^{2}}{\left(1+{\sqrt{2}} \ ma\right)}\right],
\end{eqnarray}
where the first term in square brackets on the right-hand side corresponds to the conventional Yukawa force, while the second term accounts for the contribution induced by the non-local component of the model.

\begin{figure}[!h]
\centering \includegraphics[scale=0.45]{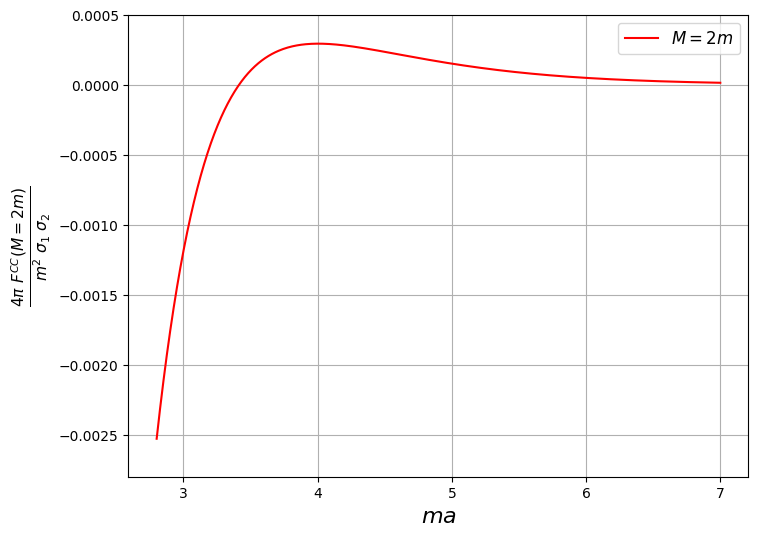} \caption{Interaction force from Eq.~(\ref{FCCS2}), multiplied by the factor $\frac{4\pi}{\sigma_1 \sigma_2 m^2}$, as a function of the dimensionless parameter $ma$, for $M = 2m$. The force is negative for small $ma$, becomes positive near $ma \approx 3.3$, and reaches a maximum around $ma \approx 4$, before gradually decreasing. This behavior confirms the transition from attractive to repulsive interaction observed in the corresponding energy profile of Fig.~\ref{grafico3}.}
\label{grafico4}
\end{figure}

Figure (\ref{grafico4}) shows a plot of the interaction force (\ref{FCCS2}) as a function of $ma$, which illustrates its overall behavior. We can see that the interaction force (\ref{FCCS2}) is attractive for values of $ma<2+{\sqrt{2}}$, with its magnitude decreasing as $ma$ increases. For small values of $ma$, the force (\ref{FCCS2}) exhibits a Yukawa-like behavior. At $ma=2+{\sqrt{2}}$, the interaction force given by Eq. (\ref{FCCS2}) vanishes, implying that the Yukawa term and the contribution arising from non-locality exactly cancel each other. Moreover, for $ma>2+{\sqrt{2}}$, the interaction force becomes repulsive, as the contribution arising from the model's non-locality dominates over the Yukawa term. For large values of $ma$ ($ma \gg 2+{\sqrt{2}}$), we obtain a long-range interaction force. We also observe that, starting from $ma=2+{\sqrt{2}}$, the force (\ref{FCCS2}) increases with $ma$ until it reaches a maximum at $ma \approx 4$. Beyond this point, the force decreases as $ma$ increases. In the limit $ma\rightarrow\infty$, the interaction force (\ref{FCCS2}) tends to zero.

In the third case, we consider the configuration for which $0<{4m^{2}}/{M^{2}}<1$. In this scenario, we have two field modes with different non-vanishing masses, $m_{+}$ and $m_{-}$, both of them smaller than $M$ and greater than $m$. In this setup, the propagator can be rewritten in the form
\begin{eqnarray}
\label{prop43}   
D\left(x,y\right)\mid_{m_{+}, m_{-}}=-\frac{1}{\left(m_{+}^{2}-m_{-}^{2}\right)}\int\frac{d^{4} p}{\left(2\pi\right)^{4}}\left(\frac{m_{+}^{2}-m^{2}}{p^{2}-m_{+}^{2}}-\frac{m_{-}^{2}-m^{2}}{p^{2}-m_{-}^{2}}\right)e^{-ip\cdot\left(x-y\right)} \ .
\end{eqnarray}

Using the same procedures employed previously, we find the following result for the interaction energy between the two scalar charges:
\begin{eqnarray}
\label{ECCS6}   
E^{CC}\left(m_{+},m_{-}\right)=-\frac{\sigma_{1}\sigma_{2}}{4\pi}\frac{1}{\left(m_{+}^{2}-m_{-}^{2}\right)}\left[\left(m_{+}^{2}-m^{2}\right)\frac{e^{-m_{+} \ a}}{a}-\left(m_{-}^{2}-m^{2}\right)\frac{e^{-m_{-} \ a}}{a}\right] \ .
\end{eqnarray}

To better understand the behavior of the energy in Eq. (\ref{ECCS6}), it is convenient to rewrite it in terms of the dimensionless quantities \( m'_{+} \) and \( m'_{-} \), defined by
\begin{eqnarray}
\label{m+m-li}    
m'_{+}=\frac{m_{+}}{m}    \  \ ; \ \  m'_{-}=\frac{m_{-}}{m} \ ,
\end{eqnarray}
In this case, we obtain
\begin{eqnarray}
 \label{ECCS7}
 E^{CC}\left(m'_{+},m'_{-}\right)&=&-\frac{\sigma_{1}\sigma_{2}}{4\pi}\frac{m}{\left[(m'_{+})^{2}-(m'_{-})^{2}\right]}\Biggl[\left[(m'_{+})^{2}-1\right]\frac{e^{-\left({m'_{+}}\right) \ ma}}{ma}\nonumber\\
&
&-\left[(m'_{-})^{2}-1\right]\frac{e^{-\left({m'_{-}}\right) \ ma}}{ma}\Biggr] \ .
\end{eqnarray}

In Fig. (\ref{FIG5}), we present a plot of the energy (\ref{ECCS7}) multiplied by $\frac{4\pi}{m\sigma_{1}\sigma_{2}}$ as a function of $ma$, for different regimes, considering four cases: the blue line represents $M=\sqrt{5} \ m$, the orange line corresponds to $M=\sqrt{15} \ m$, the green line represents $M=\sqrt{50} \ m$, and the red line corresponds to $M=\sqrt{100} \ m$. 

\begin{figure}[H]
  \centering
  \begin{subfigure}[t]{0.48\textwidth}
    \centering
    \includegraphics[width=\linewidth]{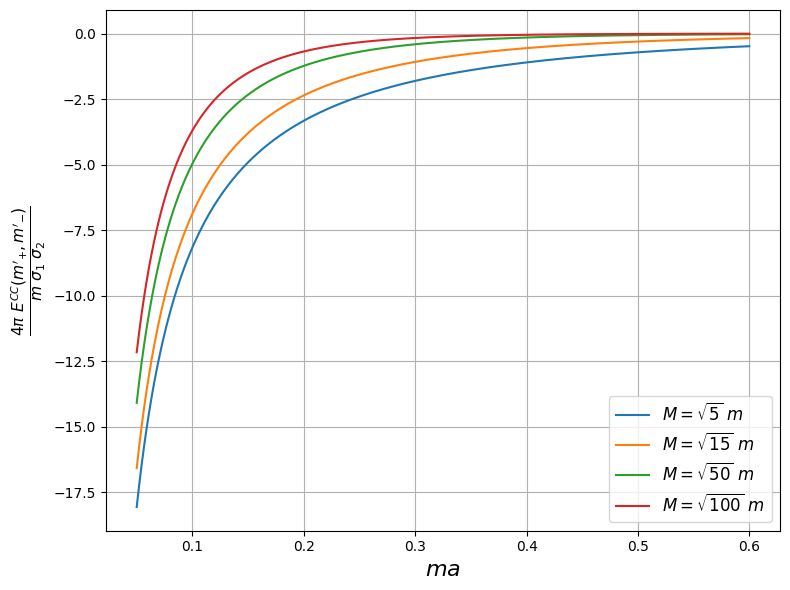}
    \caption{Small values of $ma$.}
    \label{fig:5}
  \end{subfigure}
  \hfill
  \begin{subfigure}[t]{0.48\textwidth}
    \centering
    \includegraphics[width=\linewidth]{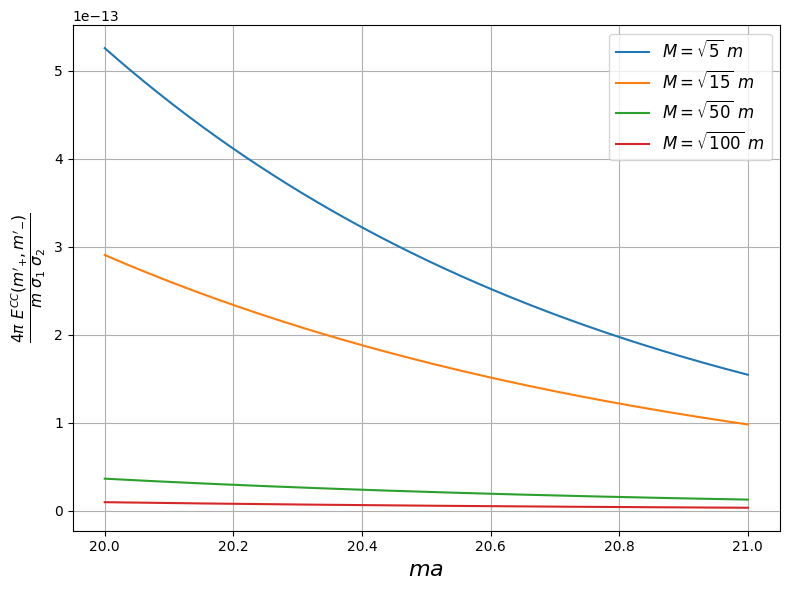}
    \caption{Interval: $20\leq ma \leq 21$.}
    \label{fig:6}
  \end{subfigure}
\hfill
\begin{subfigure}[t]{0.48\textwidth}
    \centering
    \includegraphics[width=\linewidth]{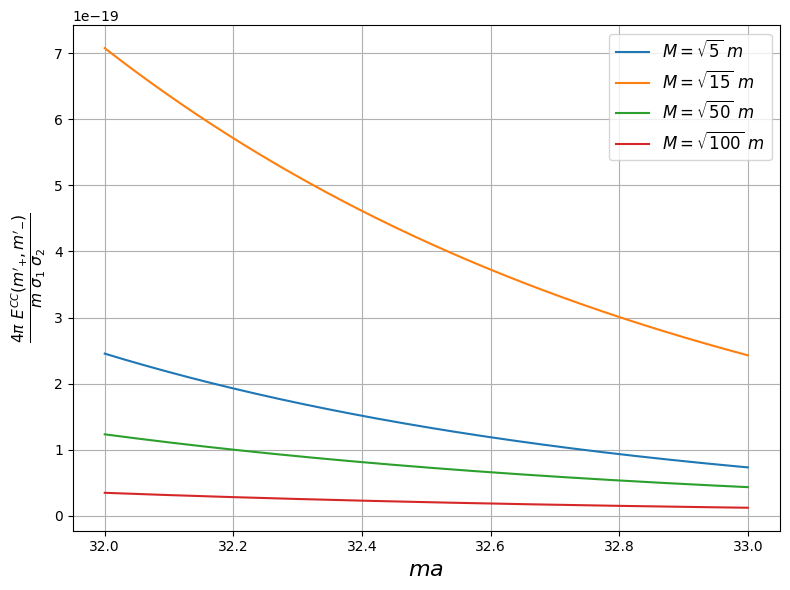}
    \caption{Interval: $32\leq ma \leq 33$.}
    \label{fig:7}
  \end{subfigure}
\hfill
\begin{subfigure}[t]{0.48\textwidth}
    \centering
    \includegraphics[width=\linewidth]{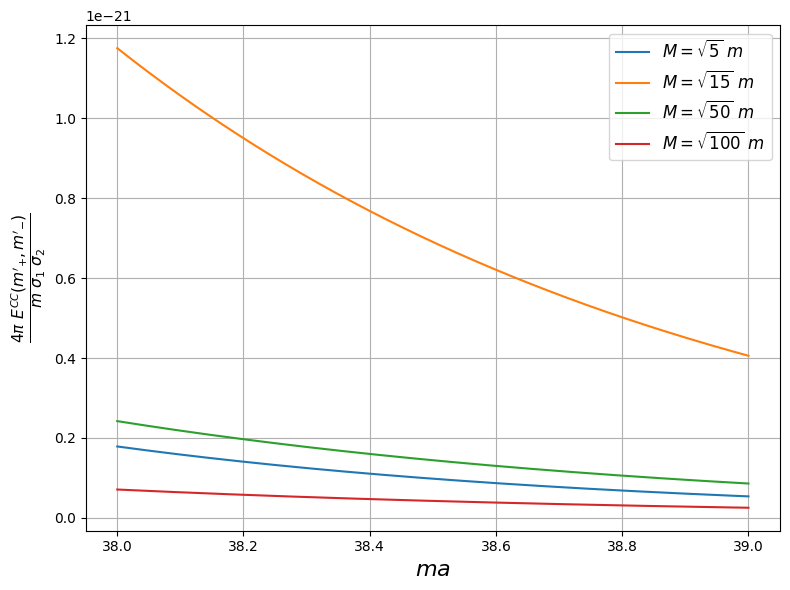}
    \caption{Interval: $38\leq ma \leq 39$.}
    \label{fig:8}
  \end{subfigure}
\hfill
\begin{subfigure}[t]{0.48\textwidth}
    \centering
    \includegraphics[width=\linewidth]{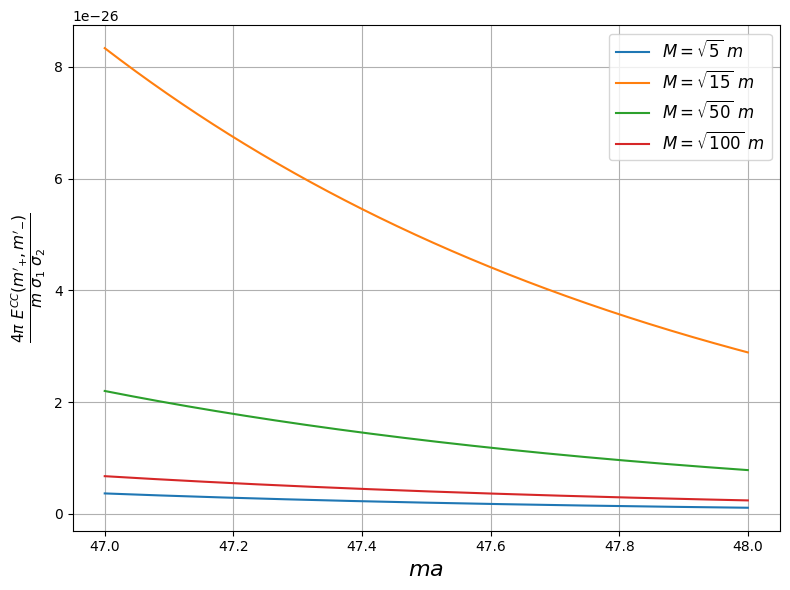}
    \caption{interval: $47\leq ma \leq 48$.}
    \label{fig:9}
  \end{subfigure}
\hfill
\begin{subfigure}[t]{0.48\textwidth}
    \centering
    \includegraphics[width=\linewidth]{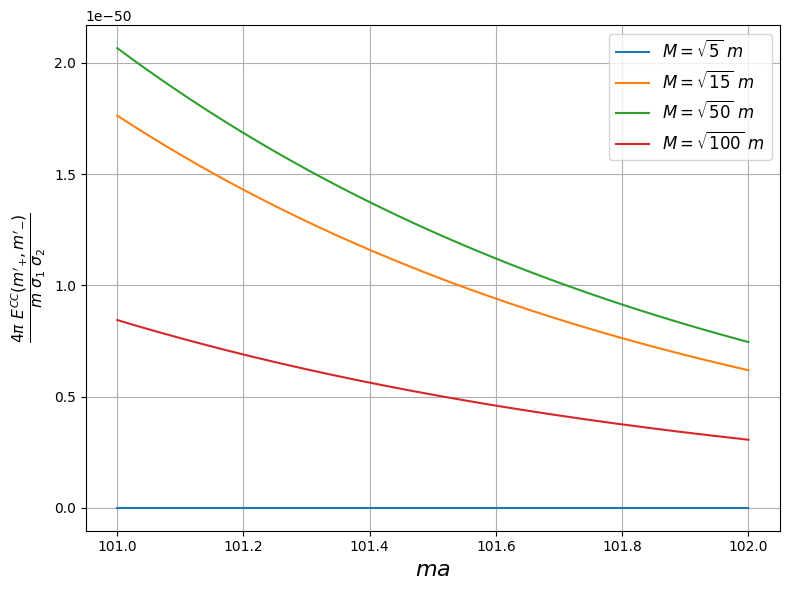}
    \caption{Interval: $101\leq ma \leq 102$.}
    \label{fig:10}
  \end{subfigure}
\end{figure}

\begin{figure}[H]
\centering
\ContinuedFloat
    \centering
    \begin{subfigure}[t]{0.48\textwidth}
    \includegraphics[width=\linewidth]{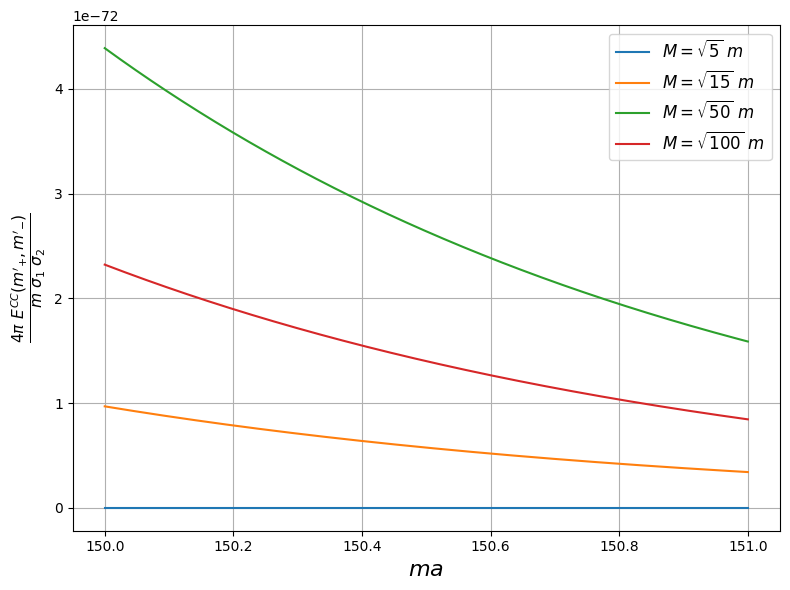}
    \caption{Interval: $150\leq ma \leq 151$.}
    \label{fig:11}
  \end{subfigure}
\hfill
\begin{subfigure}[t]{0.48\textwidth}
    \centering
    \includegraphics[width=\linewidth]{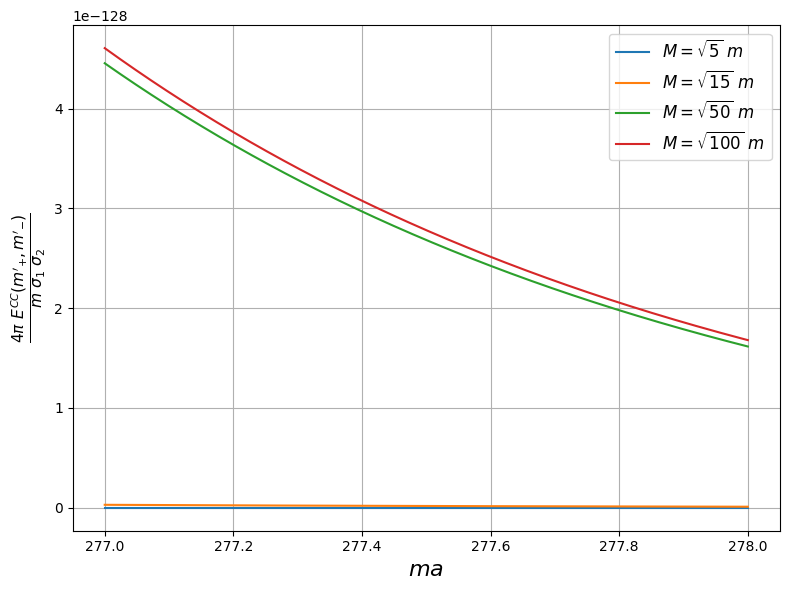}
    \caption{Interval: $277\leq ma \leq 278$.}
    \label{fig:12}
  \end{subfigure}
\hfill
\begin{subfigure}[t]{0.48\textwidth}
    \centering
    \includegraphics[width=\linewidth]{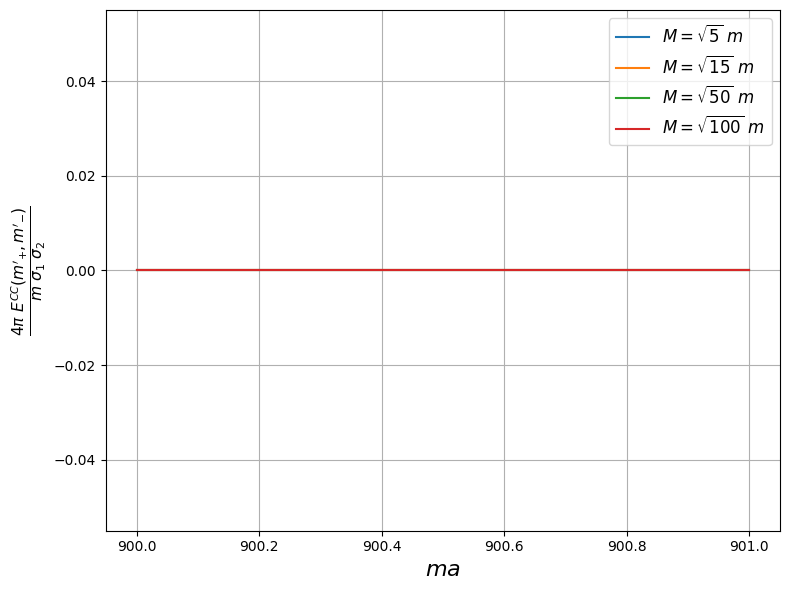}
    \caption{Interval: $900\leq ma \leq 901$.}
    \label{fig:13}
  \end{subfigure}\caption{Interaction energy from Eq.~(\ref{ECCS7}), multiplied by the factor $\frac{4\pi}{m\sigma_1\sigma_2}$, as a function of the dimensionless parameter $ma$, for different values of $M$. Figure (a) shows the strong attractive interaction at small $ma$, where the energy is negative and depends significantly on $M$. Figures (b)–(h) explore progressively larger intervals of $ma$, revealing a monotonic decay of the interaction energy, which becomes increasingly suppressed. In panel (i), for $ma \gg 1$, the energy approaches zero, indicating the vanishing of the interaction at large distances. A change in the hierarchy between the curves is observed as $ma$ increases: for small $ma$, smaller values of $M$ yield deeper energies, while at large $ma$ the curves tend to invert this ordering before converging. In this latter case, the interaction energy shifts to the repulsive regime.}\label{FIG5}
\end{figure}

By analyzing Fig.~(\ref{FIG5}), we can describe the general behavior of the interaction energy (\ref{ECCS7}). For small values of $ma$, it is always negative, monotonic, and divergent at $a=0$, as shown in Fig.~(\ref{fig:5}). In contrast to the ordinary Yukawa interaction, however, it becomes positive for a finite value of $ma$ and reaches a local maximum at another value of $ma$. For larger values of $ma$, the interaction energy (\ref{ECCS7}) remains positive and approaches zero as $ma$ increases, as illustrated in Figs.~(\ref{fig:6})–(\ref{fig:13}). 

To analyze more closely the effects of nonlocality introduced by the parameter $m$, let us examine the dependence on $ma$. For small $ma$, the absolute value of the energy (\ref{ECCS7}) decreases with increasing mass $M$ (Fig.~\ref{fig:5}), and this tendency persists even when the energy becomes positive (Fig.~\ref{fig:6}). At intermediate values of $ma$, however, this behavior reverses: the curves exchange their relative positions as shown in Figs.~\ref{fig:7}–\ref{fig:11}. For sufficiently large $ma$, the ordering of the curves stabilizes, and from Fig.~\ref{fig:12} we see that, at fixed $ma$, the interaction energy grows with $M$. Finally, in the limit of very large $ma$, the energy approaches zero (Fig.~\ref{fig:13}).

The interaction force reads
\begin{eqnarray}
\label{FCCS6}   
F^{CC}\left(m_{+},m_{-}\right)&=&-\frac{\sigma_{1}\sigma_{2}}{4\pi}\frac{m^{2}}{\left(m_{+}^{2}-m_{-}^{2}\right)}\Biggl[\left(m_{+}^{2}-m^{2}\right)\left(1+m_{+}a\right)\frac{e^{-m_{+} \ a}}{a^{2}}\nonumber\\
&
&-\left(m_{-}^{2}-m^{2}\right)\left(1+m_{-}a\right)\frac{e^{-m_{-} \ a}}{a^{2}}\Biggr] \nonumber\\
= F^{CC}\left(m'_{+},m'_{-}\right)&=&-\frac{\sigma_{1}\sigma_{2}}{4\pi}\frac{m^{2}}{\left[(m'_{+})^{2}-(m'_{-})^{2}\right]}\Biggl\{\left[(m'_{+})^{2}-1\right]\left[1+(m'_{+}) ma\right]\frac{e^{-\left({m'_{+}}\right) \ ma}}{\left(ma\right)^{2}}\nonumber\\
&
&-\left[(m'_{-})^{2}-1\right]\left[1+({m'_{-}}) ma\right]\frac{e^{-\left({m'_{-}}\right) \ ma}}{\left(ma\right)^{2}}\Biggr\} \ .
\end{eqnarray}

Figure (\ref{FIG6}) displays the behavior of the interaction force (\ref{FCCS6}), scaled by the factor $\frac{4\pi}{m^{2}\sigma_{1}\sigma_{2}}$, plotted as a function of $ma$ over different regimes — the same ones previously considered for the interaction energy (\ref{ECCS7}).
\begin{figure}[H]
  \centering
    \begin{subfigure}[t]{0.48\textwidth}
    \centering
    \includegraphics[width=\linewidth]{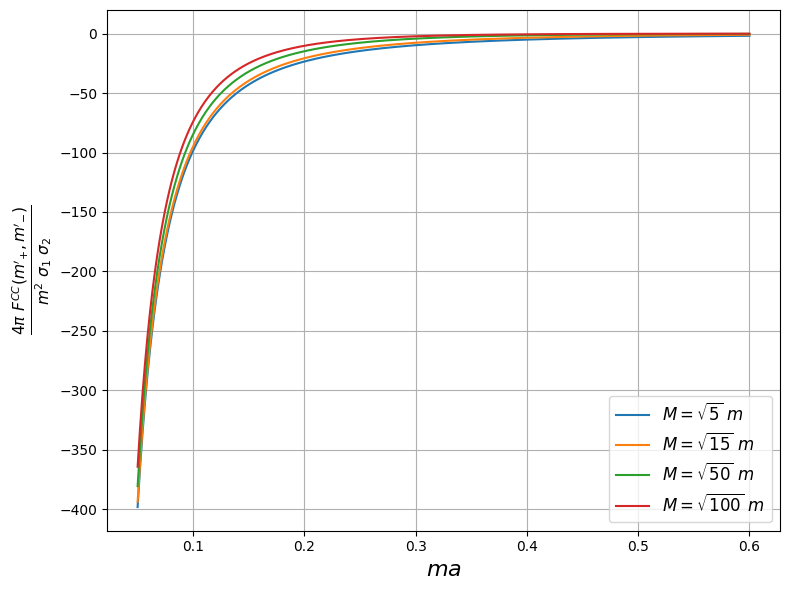}
    \caption{Small values of $ma$.}
    \label{fig:14}
  \end{subfigure}
  \hfill
  \begin{subfigure}[t]{0.48\textwidth}
    \centering
    \includegraphics[width=\linewidth]{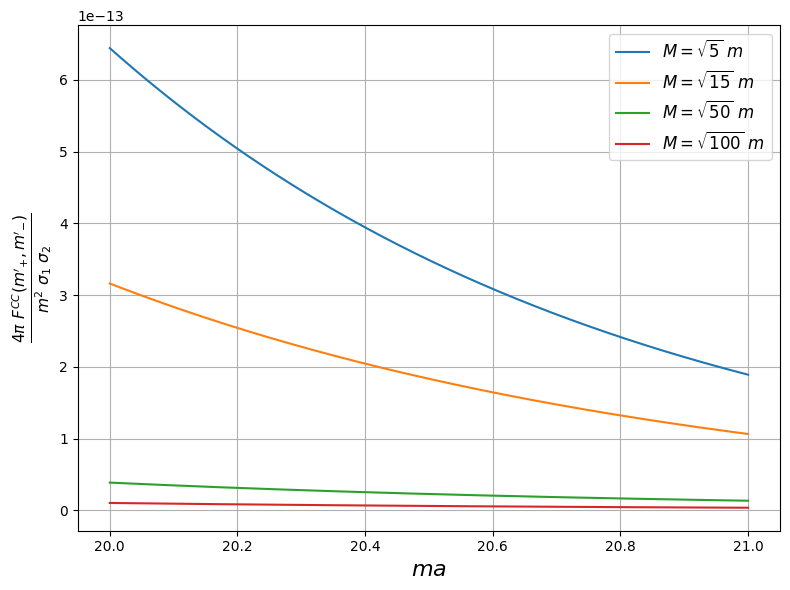}
    \caption{Interval: $20\leq ma \leq 21$.}
    \label{fig:15}
  \end{subfigure}
\hfill
\begin{subfigure}[t]{0.48\textwidth}
    \centering
    \includegraphics[width=\linewidth]{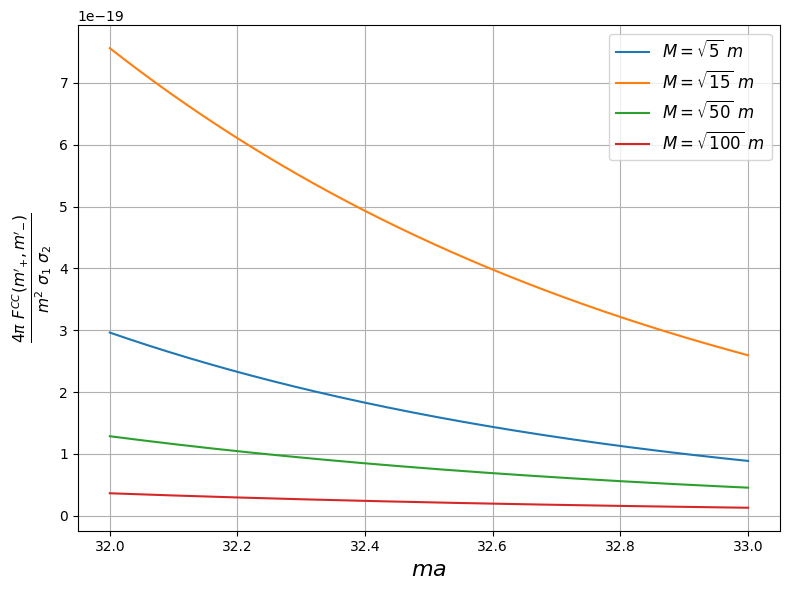}
    \caption{Interval: $32\leq ma \leq 33$.}
    \label{fig:16}
  \end{subfigure}
\hfill
\begin{subfigure}[t]{0.48\textwidth}
    \centering
    \includegraphics[width=\linewidth]{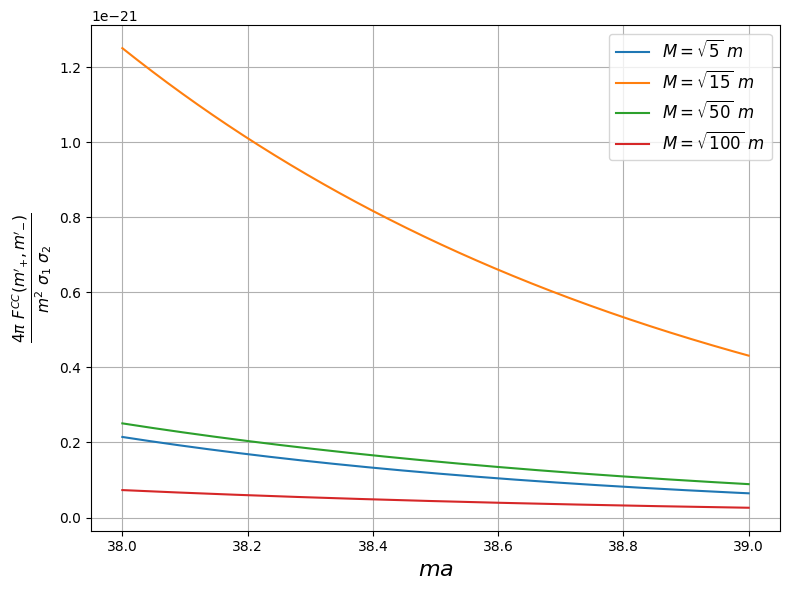}
    \caption{Interval: $38\leq ma \leq 39$.}
    \label{fig:17}
  \end{subfigure}
\hfill
\begin{subfigure}[t]{0.48\textwidth}
    \centering
    \includegraphics[width=\linewidth]{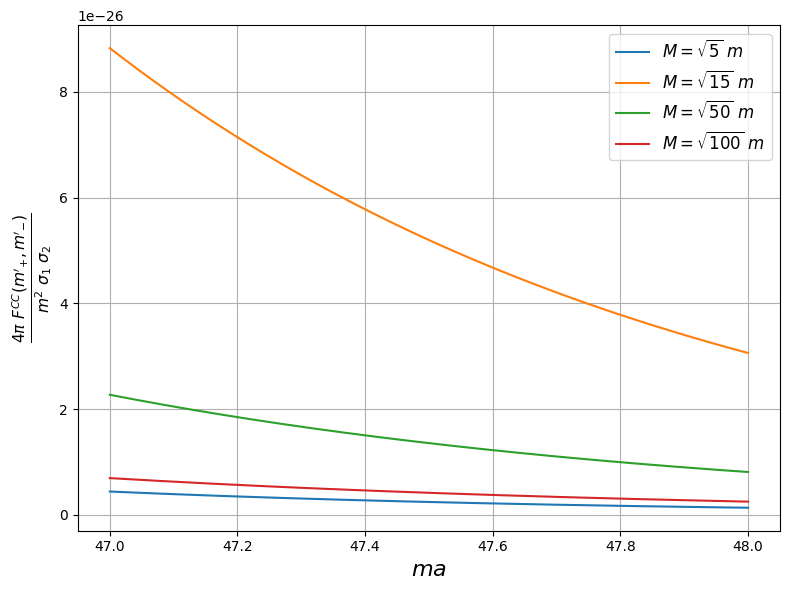}
    \caption{interval: $47\leq ma \leq 48$.}
    \label{fig:18}
  \end{subfigure}
\hfill
\begin{subfigure}[t]{0.48\textwidth}
    \centering
    \includegraphics[width=\linewidth]{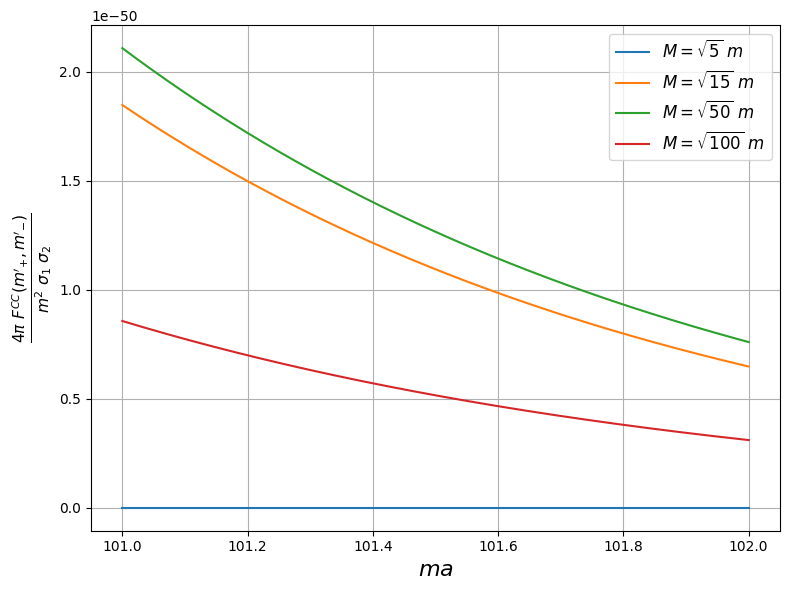}
    \caption{Interval: $101\leq ma \leq 102$.}
    \label{fig:19}
  \end{subfigure}
\end{figure}

\begin{figure}[H]
\centering
\ContinuedFloat
    \centering
    \begin{subfigure}[t]{0.48\textwidth}
    \includegraphics[width=\linewidth]{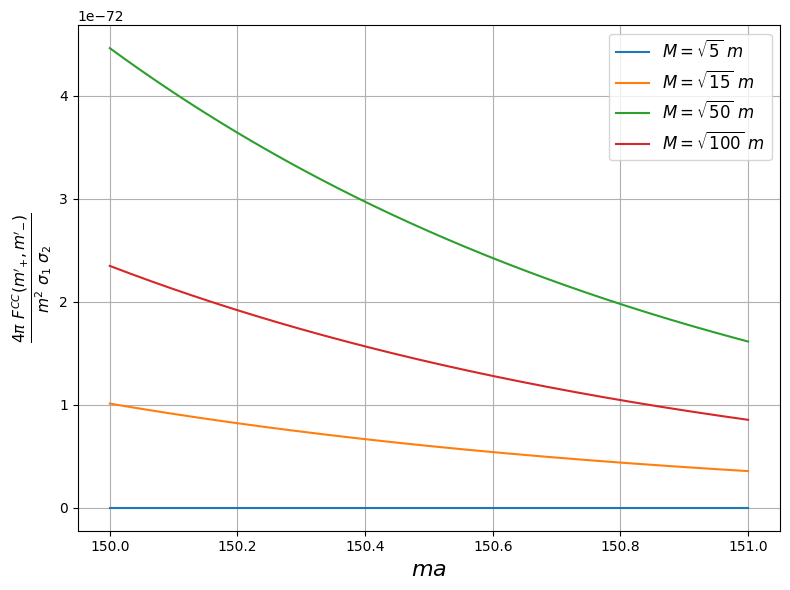}
    \caption{Interval: $150\leq ma \leq 151$.}
    \label{fig:20}
  \end{subfigure}
\hfill
\begin{subfigure}[t]{0.48\textwidth}
    \centering
    \includegraphics[width=\linewidth]{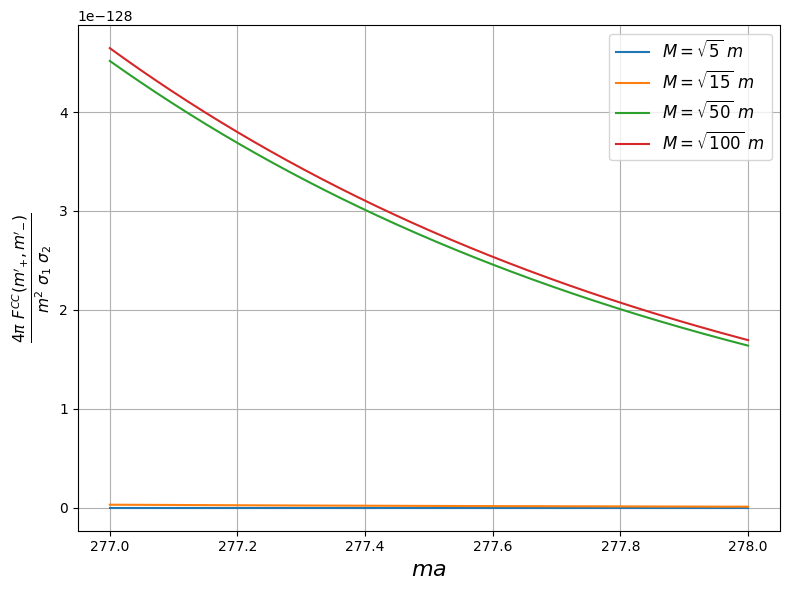}
    \caption{Interval: $277\leq ma \leq 278$.}
    \label{fig:21}
  \end{subfigure}
\hfill
\begin{subfigure}[t]{0.48\textwidth}
    \centering
    \includegraphics[width=\linewidth]{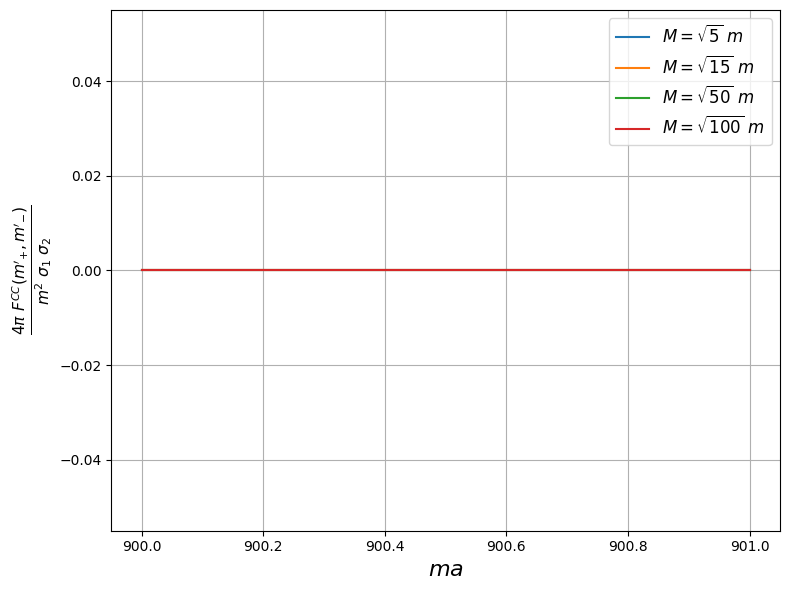}
    \caption{Interval: $900\leq ma \leq 901$.}
    \label{fig:22}
  \end{subfigure}
  \caption{Interaction force from Eq.~(\ref{FCCS6}), multiplied by the factor $\frac{4\pi}{m^2 \sigma_1 \sigma_2}$, as a function of the dimensionless parameter $ma$, for various values of $M$. Figure (a) shows strong attractive forces for small $ma$, with magnitude decreasing rapidly as $ma$ increases. Figuress (b)–(h) illustrate the asymptotic suppression of the force in extended intervals, revealing a monotonic decay with $ma$. In Figure (i), for $ma \gg 1$, the force vanishes numerically, confirming the disappearance of the interaction at large distances. A change in the hierarchy between the curves is also observed: for small $ma$, smaller $M$ yields stronger forces, while for larger $ma$, the relative ordering between curves inverts before converging. In this latter case, the force shifts to the repulsive regime.}\label{FIG6}
\end{figure}

From Fig.~\ref{FIG6}, we observe that for small values of $ma$, the interaction force (\ref{FCCS6}) diverges at $a=0$, is attractive for equal signs of $\sigma_{1}$ and $\sigma_{2}$, and decreases in magnitude as $ma$ increases, as illustrated in Fig.~\ref{fig:14}. For larger values of $ma$, the force grows again, vanishes at a finite point, and then becomes positive (for equal signs of $\sigma_{1}$ and $\sigma_{2}$). It reaches a local maximum at another finite value of $ma$, after which it remains positive and gradually approaches zero in the limit $ma \to \infty$. Thus, for sufficiently large $ma$, where nonlocal effects dominate, the interaction force becomes repulsive, as shown in Figs.~\ref{fig:15}–\ref{fig:22}.

In conventional scalar field theory (i.e., without the nonlocal term), the smaller the mass $M$, the larger the magnitude of the interaction force for fixed $ma$ (Fig.~\ref{fig:14}). In the model (\ref{model1}), this behavior is preserved only for small $ma$, but gradually reverses as $ma$ increases, as illustrated in Figs.~\ref{fig:15}–\ref{fig:20}. By the time we reach Fig.~\ref{fig:21}, the trend has inverted: the greater the mass $M$, the stronger the force (\ref{FCCS6}). From that point onward, this behavior persists, with the force remaining positive and tending to zero as $ma$ grows, as shown in Fig.~\ref{fig:22}.

\section{The propagator in the presence of a Dirichlet plane}
\label{III}

In this section, we analyze the field governed by the non-local model (\ref{model1}) in the presence of a single Dirichlet plane. The presence of this plane enforces the condition that the field must vanish on it. Without loss of generality, we adopt a coordinate system in which the Dirichlet plane is oriented perpendicular to the $x^{3}$ axis and located at $x^{3}=a$. Accordingly, the Dirichlet boundary condition imposed on the field can be written as
\begin{eqnarray}
\label{Dirich}
\phi\left(x\right)\mid_{x^{3}=a}=0 \ ,
\end{eqnarray}
where the subscript indicates that the boundary condition is taken on the plane $x^{3}=a$.

Our goal is to determine the generating functional for the scalar field under the boundary condition (\ref{Dirich}). To accomplish this, we adopt the functional formalism introduced in \cite{BRW} and employed in \cite{LHCBFABplate,BC14,BC15,BC16,F8,F9}. We begin by expressing the generating functional as follows:
\begin{eqnarray}
\label{fgen1}
Z_{C}\left[J\right]=\int {\cal{D}}\phi_{C} \ e^{i\int d^{4}x \ \cal{L}} \ ,
\end{eqnarray}
Here, the subscript $C$ indicates that the integration is carried out over all field configurations that satisfy the condition (\ref{Dirich}). This constraint is implemented by introducing a functional delta, which is non-zero only when the condition (\ref{Dirich}) is fulfilled, as defined below:
\begin{eqnarray}
\label{fgen2}
Z_{C}\left[J\right]=\int {\cal{D}}\phi \ \delta\left[\phi\left(x\right)\mid_{x^{3}=a}\right] \ e^{i\int d^{4}x \ \cal{L}} \ .
\end{eqnarray}

The delta functional appearing in (\ref{fgen2}) can be expressed in terms of its Fourier representation, as follows:
\begin{eqnarray}
\label{fgen3}
\delta\left[\phi\left(x\right)\mid_{x^{3}=a}\right] =\int {\cal{D}}B\exp\left[-i
\int d^{4}x\ \delta\left(x^{3}-a\right)B\left(x_{\parallel}\right)\phi
\left(x\right)\right] \ ,
\end{eqnarray}
where $B\left(x_{\parallel}\right)$ is an auxiliary scalar field and $x_{\parallel}^{\mu}=\left(x^{0},x^{1},x^{2}\right)$ indicates that we consider only the coordinates parallel to the plane.

Substituting Eq.~(\ref{fgen3}) into (\ref{fgen2}), we arrive at
\begin{eqnarray}
\label{fgen4}
Z_{C}\left[J\right]=\int {\cal{D}}\phi {\cal{D}}B\ e^{i\int d^{4}x \ \cal{L}}\exp\left[-i
\int d^{4}x\ \delta\left(x^{3}-a\right)B\left(x_{\parallel}\right)\phi
\left(x\right)\right] \ .
\end{eqnarray}

To cast the above integral into a more convenient form, we perform the field shift:
\begin{eqnarray}
\label{Translation}
\phi\left(x\right)\rightarrow\phi\left(x\right)-\int d^{4}y \ \delta\left(y^{3}-a\right)D\left(x,y\right)B\left(y_{\parallel}\right) \ ,
\end{eqnarray}
which has a unit Jacobian. This transformation allows us to rewrite Eq.~(\ref{fgen4}) as
\begin{eqnarray}
\label{fgen5}
Z_{C}\left[J\right]=Z\left[J\right]{\bar{Z}}\left[J\right] \ ,
\end{eqnarray}
where $Z\left[J\right]$ is the standard generating functional for the scalar field,
\begin{eqnarray}
\label{fgen6}
Z\left[J\right]=\int{\cal{D}}\phi\ e^{i\int d^{4}x \ \cal{L}}
=Z\left[0\right]\exp\left[\frac{i}{2}\int d^{4}x \ d^{4}y \ J\left(x\right)D\left(x,y\right)J\left(y\right)\right] \ ,
\end{eqnarray}
and ${\bar{Z}}\left[J\right]$ is the contribution arising from the auxiliary scalar field $B$,
\begin{eqnarray}
\label{fgen7}
{\bar{Z}}\left[J\right]=\int{\cal{D}}B\exp\left[-i\int d^{4}y \ \delta
\left(y^{3}-a\right)I\left(y\right)B\left(y_{\parallel}\right)\right] \nonumber\\
\times\exp\left[\frac{i}{2}\int d^{4}x \ d^{4}y \ \delta\left(x^{3}-a\right)
\delta\left(y^{3}-a\right)B\left(x_{\parallel}\right)D\left(x,y\right)
B\left(y_{\parallel}\right)\right] \ ,
\end{eqnarray}
where we have defined
\begin{eqnarray}
\label{defi1}
I\left(y\right)=\int d^{4}x \ D\left(x,y\right)J\left(x\right)  \ . 
\end{eqnarray}

From now on, we will consider three types of configurations separately, as we did in the previous section, namely: $M=0$, $M=2m$ and $0<4m^2/M^2<1$.

In the first case, let us consider the situation where $M=0$. Substituting Eqs.~(\ref{defi1}) and (\ref{prop31}) into Eq.~(\ref{fgen7}), and using the identities
\begin{eqnarray}
\label{intp31}
\int \frac{dp^{3}}{2\pi}\frac{e^{i p^{3}(x^{3}-y^{3})}}{p^{2}} &=& -\frac{i}{2\Gamma} e^{i\Gamma\mid x^{3}-y^{3}\mid} \ , \\
\label{intp312}
\int \frac{dp^{3}}{2\pi}\frac{e^{i p^{3}(x^{3}-y^{3})}}{p^{4}} &=& -\frac{1}{4 p_{\parallel}^{2}}\left(\frac{i}{\Gamma}+\mid x^{3}-y^{3}\mid\right) e^{i\Gamma\mid x^{3}-y^{3}\mid} \ ,
\end{eqnarray}
where $p^{3}$ is the momentum perpendicular to the plane, $\Gamma=\sqrt{p_{\parallel}^{2}}$, and
$p_{\parallel}^{\mu}=\left(p^{0},p^{1},p^{2}\right)$ denotes the momentum parallel to the plane, we obtain
\begin{eqnarray}
\label{ZJpara1}
{\bar{Z}}\left[J\right] &=& \int {\cal{D}}B \exp\left[-i\int d^{3}y_{\parallel} \ I\left(y_{\parallel}\right)B\left(y_{\parallel}\right)\right] \nonumber\\
&\times& \exp\left[\frac{i}{2}\int d^{3}x_{\parallel} \ d^{3}y_{\parallel} \ B\left(x_{\parallel}\right)D\left(x_{\parallel},y_{\parallel}\right)B\left(y_{\parallel}\right)\right] \ ,
\end{eqnarray}
where
\begin{eqnarray}
\label{Dxypara}
D\left(x_{\parallel},y_{\parallel}\right)=\frac{i}{2}\int\frac{d^{3}p_{\parallel}}{\left(2\pi\right)^{3}} \ e^{-ip_{\parallel}\cdot\left(x_{\parallel}-y_{\parallel}\right)}\left(1-\frac{m^{2}}{2p_{\parallel}^{2}}\right)\frac{1}{\Gamma} \ ,
\end{eqnarray}
\begin{eqnarray}
\label{Iypara}
I\left(y_{\parallel}\right)=\int d^{4}x \  J\left(x\right)f\left(x,y_{\parallel}\right) \ ,
\end{eqnarray}
with the function
\begin{eqnarray}
\label{fxypara}
f\left(x,y_{\parallel}\right)=\frac{i}{2}\int\frac{d^{3}p_{\parallel}}{\left(2\pi\right)^{3}} \ e^{-ip_{\parallel}\cdot\left(x_{\parallel}-y_{\parallel}\right)}\left[1+\frac{im^{2}}{2\Gamma}\left(\frac{i}{\Gamma}+\mid x^{3}-a\mid\right)\right]\frac{e^{i\Gamma\mid x^{3}-a\mid}}{\Gamma} \ .
\end{eqnarray}

Now, in the functional integral (\ref{ZJpara1}), we perform the following translation:
\begin{eqnarray}
\label{traBpara}
B\left(x_{\parallel}\right)\rightarrow B\left(x_{\parallel}\right)+\int d^{3}y_{\parallel}V\left(x_{\parallel},y_{\parallel}\right)
I\left(y_{\parallel}\right) \ ,
\end{eqnarray}
where $V\left(x_{\parallel},y_{\parallel}\right)$ is the inverse of
$D\left(x_{\parallel},y_{\parallel}\right)$, that is,
\begin{eqnarray}
\label{invedxypara}
\int d^{3}y_{\parallel}D\left(x_{\parallel},y_{\parallel}\right)V\left(y_{\parallel},z_{\parallel}\right) =\delta^{3}\left(x_{\parallel}-z_{\parallel}\right) \ ,
\end{eqnarray}
with
\begin{eqnarray}
\label{Vyzpara}
V\left(y_{\parallel},z_{\parallel}\right)=-4i\int\frac{d^{3}p_{\parallel}}{\left(2\pi\right)^{3}} \ e^{-ip_{\parallel}\cdot\left(y_{\parallel}-z_{\parallel}\right)}\frac{\Gamma \  p_{\parallel}^{2}}{\left(2p_{\parallel}^{2}-m^{2}\right)} \ .
\end{eqnarray}
This leads to
\begin{eqnarray}
\label{fgen8}
{\bar{Z}}\left[J\right]={\bar{Z}}\left[0\right]\exp\left[\frac{i}{2}\int d^{4}x 
\ d^{4}y \ J\left(x\right){\bar{D}}\left(x,y\right)J\left(y\right)\right] \ ,
\end{eqnarray}
where we define
\begin{eqnarray}
\label{fgfvf}
{\bar{D}}\left(x,y\right)=-\int d^{3}\omega_{\parallel} \ d^{3}z_{\parallel} \ f\left(x,\omega_{\parallel}\right)V\left(\omega_{\parallel},z_{\parallel}\right)f\left(y,z_{\parallel}\right) \ .
\end{eqnarray}

Substituting Eqs.~(\ref{fxypara}) and (\ref{Vyzpara}) into Eq.~(\ref{fgfvf}) and carrying out the necessary calculations, we find
\begin{eqnarray}
\label{prpoplate2}
{\bar{D}}_{\mu\nu}\left(x,y\right)\mid_{M=0}&=&-\frac{i}{2}\int\frac{d^{3}p_{\parallel}}{\left(2\pi\right)^{3}} \ e^{-ip_{\parallel}\cdot\left(x_{\parallel}-y_{\parallel}\right)}
\frac{2\Gamma}{\left(2p_{\parallel}^{2}-m^{2}\right)}\left[1+\frac{im^{2}}{2\Gamma}\left(\frac{i}{\Gamma}+\mid x^{3}-a\mid\right)\right]\nonumber\\
&
&\times\left[1+\frac{im^{2}}{2\Gamma}\left(\frac{i}{\Gamma}+\mid y^{3}-a\mid\right)\right]e^{i\Gamma\left(\mid x^{3}-a\mid +\mid y^{3}-a\mid\right)} \ .
\end{eqnarray}

From Eq.~(\ref{fgen6}) and using Eq.~(\ref{fgen8}), the functional generator in Eq.~(\ref{fgen5}) becomes
\begin{eqnarray}
\label{fgen9}
Z_{C}\left[J\right]=Z_{C}\left[0\right]\exp\left[\frac{i}{2}
\int d^{4}x \ d^{4}y \ J\left(x\right)\left(D
\left(x,y\right)\mid_{M=0}+{\bar{D}}\left(x,y\right)\mid_{M=0}\right)J
\left(y\right)\right] \ .
\end{eqnarray}

With Eq.~(\ref{fgen9}) at hand, we identify the propagator in the presence of a Dirichlet plane as
\begin{eqnarray}
\label{Proptotal}
D_{C}\left(x,y\right)\mid_{M=0}=D\left(x,y\right)\mid_{M=0}+{\bar{D}}\left(x,y\right)\mid_{M=0} \ .
\end{eqnarray}

 The propagator in Eq.~(\ref{Proptotal}) is given by the free contribution (\ref{prop31}) plus the correction term (\ref{prpoplate2}), which accounts for the effects introduced by the Dirichlet plane. In the limit $m \rightarrow 0$, the correction term (\ref{prpoplate2}) reduces to the expression found in the standard massless scalar field theory with Dirichlet boundary conditions  \cite{RBEFAnderson}.

To clarify the actual role of the Dirichlet boundary condition in this result, we analyze the classical field solution in the presence of an external source:
\begin{eqnarray}
\label{Field}
\phi\left(x\right)=\int d^{4}y \  \left(D_{C}\left(x,y\right)\mid_{M=0}\right)J\left(y\right) \ .
\end{eqnarray}
Since $\left(D_{C}\left(x,y\right)\mid_{M=0}\right)\mid_{x^{3}=a}=0$, we confirm that the solution in Eq.~(\ref{Field}) satisfies the boundary condition (\ref{Dirich}).

In the next scenario, we analyze the case in which $M=2m$. Using the identity
\begin{eqnarray}
\label{}
\int \frac{dp^{3}}{2\pi}\frac{e^{i p^{3}(x^{3}-y^{3})}}{p^{2}-2m^{2}} = -\frac{i}{2L} e^{iL\mid x^{3}-y^{3}\mid} \ ,
\end{eqnarray}
where $L=\sqrt{p_{\parallel}^{2}-2m^{2}}$, and following a procedure analogous to the one used previously, we find that the correction to the propagator (\ref{prop32}) induced by the presence of the Dirichlet plane is given by
\begin{eqnarray}
\label{prpoplate3}
{\bar{D}}_{\mu\nu}\left(x,y\right)\mid_{M=2m}&=&-\frac{i}{4}\int\frac{d^{3}p_{\parallel}}{\left(2\pi\right)^{3}} \ e^{-ip_{\parallel}\cdot\left(x_{\parallel}-y_{\parallel}\right)}
\frac{1}{\left(2p_{\parallel}^{2}-3 m^{2}\right)}\left[2p_{\parallel}^{2}-m^{2}\left(3+iL\mid x^{3}-a\mid\right)\right]\nonumber\\
&&\times\left[2p_{\parallel}^{2}-m^{2}\left(3+iL\mid y^{3}-a\mid\right)\right]\frac{e^{iL\left(\mid x^{3}-a\mid +\mid y^{3}-a\mid\right)}}{L^{3}} \ .
\end{eqnarray}

Analogously to the analysis carried out in the previous case, it can be demonstrated that the propagator 
\begin{eqnarray}
\label{ProptotalM2m}
D_{C}\left(x,y\right)\mid_{M=2m}=D\left(x,y\right)\mid_{M=2m}+{\bar{D}}\left(x,y\right)\mid_{M=2m} \ ,
\end{eqnarray}
indeed satisfies the Dirichlet boundary condition (\ref{cond1}).


Now, for the configuration where $0<{4m^{2}}/{M^{2}}<1$, we take into account the identities
\begin{eqnarray}
\label{intp313}
\int \frac{dp^{3}}{2\pi}\frac{e^{i p^{3}(x^{3}-y^{3})}}{p^{2}-m_{+}^{2}} = -\frac{i}{2\Gamma_{+}} e^{i\Gamma_{+}\mid x^{3}-y^{3}\mid} \ , \\
\label{intp3123}
\int \frac{dp^{3}}{2\pi}\frac{e^{i p^{3}(x^{3}-y^{3})}}{p^{2}-m_{-}^{2}} = -\frac{i}{2L_{-}} e^{iL_{-}\mid x^{3}-y^{3}\mid} \ , 
\end{eqnarray}
where $\Gamma_{+}=\sqrt{p_{\parallel}^{2}-m_{+}^{2}}$ and $L_{-}=\sqrt{p_{\parallel}^{2}-m_{-}^{2}}$.  
By employing the same methods used in the previous cases, we arrive at the following result:
\begin{eqnarray}
\label{proplate4}   
{\bar{D}}_{\mu\nu}\left(x,y\right)\mid_{m_{+},m_{-}}&=&-\frac{i}{2}\frac{1}{\left(m_{+}^{2}-m_{-}^{2}\right)}\int\frac{d^{3}p_{\parallel}}{\left(2\pi\right)^{3}} \ e^{-ip_{\parallel}\cdot\left(x_{\parallel}-y_{\parallel}\right)}\nonumber\\
&\times&\left[\left(m_{+}^{2}-m^{2}\right)\frac{1}{\Gamma_{+}}-\left(m_{-}^{2}-m^{2}\right)\frac{1}{L_{-}}\right]^{-1}\nonumber\\
&\times&\left[\left(m_{+}^{2}-m^{2}\right)\frac{e^{i\Gamma_{+}\mid x^{3}-a\mid}}{\Gamma_{+}}-\left(m_{-}^{2}-m^{2}\right)\frac{e^{iL_{-}\mid x^{3}-a\mid}}{L_{-}}\right]\nonumber\\
&\times&\left[\left(m_{+}^{2}-m^{2}\right)\frac{e^{i\Gamma_{+}\mid y^{3}-a\mid}}{\Gamma_{+}}-\left(m_{-}^{2}-m^{2}\right)\frac{e^{iL_{-}\mid y^{3}-a\mid}}{L_{-}}\right] \ .
\end{eqnarray}

Equation~(\ref{proplate4}) provides the correction to the propagator (\ref{prop43}) arising from the presence of the Dirichlet plane. In this case, it can also be shown that the full propagator satisfies the boundary condition given in (\ref{cond1}). It is worth noting that, in the limit $m \rightarrow 0$, we have $m_{+} = M$ and $m_{-} = 0$, and the propagator (\ref{proplate4}) reduces to the one obtained from the standard Klein-Gordon theory in the presence of a single Dirichlet plane \cite{RBEFAnderson}.

\section{\label{IV} Charge--plane interaction}
In this section we consider the interaction energy between a point-like external source and the Dirichlet plane, which obtained from the integral expression \cite{GTFABFEB,LHCBAFFFAB,BC16}
\begin{eqnarray}
\label{Energy}
E^{CP}=-\frac{1}{2T}\int d^{4}x \  d^{4}y \  J\left(x\right){\bar{D}}\left(x,y\right)J\left(y\right) \ ,
\end{eqnarray}
The sub-index $CP$ means that we have the interaction energy between the charge and the Dirichlet plane. With no loss of generality, and for simplicity, we choose
a point-like scalar charge placed at position  ${\bf{b}} = \left(0, 0, b\right)$. The corresponding external source reads
\begin{eqnarray}
\label{Source}
J\left(x\right)=\sigma\delta^{3}\left({\bf{x}}
-{\bf{b}}\right) \ ,
\end{eqnarray}

For $M=0$, we substitute the expressions (\ref{Source}) and (\ref{prpoplate2}) in (\ref{Energy}), and then, perform some manipulations, obtaining
\begin{eqnarray}
 \label{EPCP}
 E^{CP}\left(M=0\right)=\frac{\sigma^{2}}{8\pi^{2}}\int d^{2}{\bf {p}}_{\parallel}\frac{e^{-2R\sqrt{{\bf {p}}_{\parallel}^{2}}}\sqrt{{\bf {p}}_{\parallel}^{2}}}{\left(2{\bf {p}}_{\parallel}^{2}+m^{2}\right)}\left[1+\frac{m^{2}}{2\sqrt{{\bf {p}}_{\parallel}^{2}}}\left(\frac{1}{\sqrt{{\bf {p}}_{\parallel}^{2}}}+R\right)\right]^{2}  \ ,
\end{eqnarray}
where $R=\mid a-b\mid$ is the distance between the plane and the charge.  This result can be simplified by using polar coordinates and integrating out in the solid angle,
\begin{eqnarray}
\label{EPC2}   
 E^{CP}\left(M=0\right)&=&\frac{\sigma^{2}}{16\pi}\Biggl[2\int_{0}^{\infty}dp \ e^{-2Rp}+2m^{2}\Biggl(R\int_{0}^{\infty}dp \ \frac{e^{-2Rp}}{p}+\frac{1}{2}\int_{0}^{\infty}dp \ \frac{e^{-2Rp}}{p^{2}}\Biggr)\nonumber\\
 &
 &+m^{4}R^{2}\int_{0}^{\infty}dp \ \frac{e^{-2Rp}}{\left(2p^{2}+m^{2}\right)}\Biggr] \ .
\end{eqnarray}

Upon ev
aluating the integrals, we find
\begin{eqnarray}
\label{EPC3}    
E^{CP}\left(M=0\right)=\frac{\sigma^{2}}{16\pi R}\left[1+\Delta_{1}\left(mR\right)\right] \ ,
\end{eqnarray}
in which the function $\Delta_{1}\left(mR\right)$ is defined  as follows
\begin{eqnarray}
\label{delta1}    
\Delta_{1}\left(mR\right)&=&-2\left(mR\right)^{2}+ \frac{\left(mR\right)^{3}}{2\sqrt{2}}\Bigl\{2 \ {\textrm{Ci}}\left(mR\sqrt{2}\right)\sin\left(mR\sqrt{2}\right)\nonumber\\
&
&+\cos\left(mR\sqrt{2}\right)\left[\pi-2 \ {\textrm{Si}}\left(mR\sqrt{2}\right)\right]\Bigr\} \ ,
\end{eqnarray}
where ${\textrm{Ci}}\left(x\right)$ and  ${\textrm{Si}}\left(x\right)$ stand for cosine and sine integral functions \cite{Arfken}, defined by 
\begin{equation}
\label{CISI}    
{\textrm{Ci}}\left(x\right)=-\int_{x}^{\infty} dt \  \frac{\cos t}{t} \ , \  \  {\textrm{Si}}\left(x\right)=\int_{0}^{x} dt \  \frac{\sin t}{t} \ .
\end{equation}

\begin{figure}[!h]
\centering 
\includegraphics[scale=0.45]{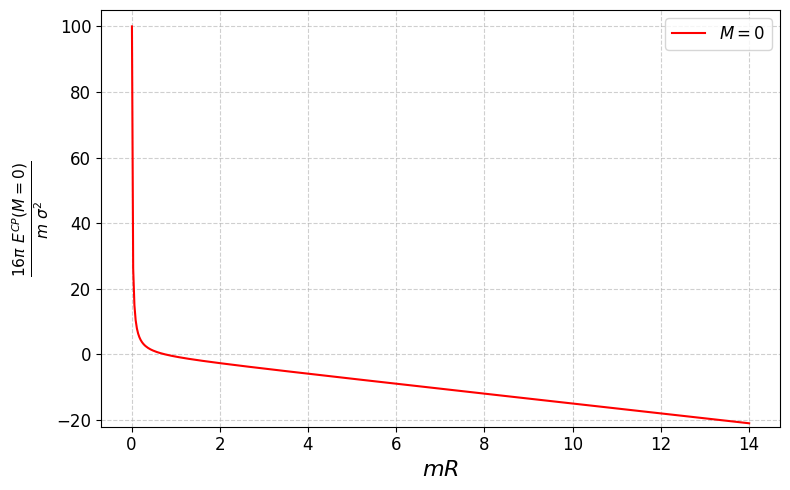} 
\caption{Interaction energy $E^{CP}$, given by (\ref{EPC3}) for $M = 0$, multiplied by the factor $\frac{16\pi}{m\sigma^2}$, as a function of the dimensionless parameter $mR$. The energy diverges positively as $mR \to 0$, crosses zero around $mR \approx 1$, and becomes negative for larger $mR$.}
\label{grafico5}
\end{figure}

The result (\ref{EPC3}) is exact and provides the interaction energy between a Dirichlet plane and a point-like scalar charge in the case where $M=0$. In Eq.~(\ref{EPC3}), the first term on the right-hand side corresponds to the standard plane-charge interaction in conventional scalar field theory. The second term represents a non-local correction introduced by the parameter $m$, encapsulated in the function $\Delta_{1}(mR)$. For non-zero values of $mR$, we have $\Delta_{1}(mR) < 0$, and its magnitude increases with increasing $mR$. Figure~\ref{grafico5} presents a plot of the energy~(\ref{EPC3}), multiplied by the factor $\frac{16\pi}{\sigma^{2}m}$, as a function of the dimensionless parameter $mR$.

From Fig.~\ref{grafico5}, we observe that in the range $0 < mR \lesssim 0.77$, the function satisfies $-1 < \Delta_{1}(mR) < 0$, which implies $E^{CP} > 0$ according to Eq.~(\ref{EPC3}), with its magnitude decreasing as $mR$ increases. For small values of $mR$, $\Delta_{1}(mR) \approx 0$, and the energy~(\ref{EPC3}) exhibits a Coulomb-like behavior, corresponding to a short-range potential. Around $mR \approx 0.77$, $\Delta_{1}(mR) \approx -1$, leading to a cancellation of the plane–charge interaction. This indicates that the non-local correction becomes comparable in magnitude to the standard Coulomb term. Finally, for $mR \gtrsim 0.77$, $\Delta_{1}(mR) < -1$, so that $E^{CP} < 0$, and its magnitude increases with $mR$. In the regime of large $mR$, where $\Delta_{1}(mR) \ll -1$, the interaction~(\ref{EPC3}) is fully dominated by the non-local contribution, leading to a long-range potential.

Equation (\ref{EPC3}) allows us to derive the interaction force between the charge and the Dirichlet plane,
\begin{eqnarray}
\label{FPC3}    
F^{CP}\left(M=0\right)=\frac{\sigma^{2}}{16\pi R^{2}}\left[1+\Delta_{2}\left(mR\right)\right] \ ,
\end{eqnarray}
in which the function $\Delta_{2}\left(mR\right)$ is given by
\begin{eqnarray}
\label{delta2}
\Delta_{2}\left(mR\right)&=&-\frac{1}{2}\Biggl\{-4\left(mR\right)^{2}+\left(mR\right)^{3}\cos\left(mR\sqrt{2}\right)\Biggl[2\left(mR\right){{\textrm{Ci}}}\left(mR\sqrt{2}\right)\nonumber\\
&
&+{\sqrt{2}}\Bigl[\pi-2 \ {\textrm{Si}}\left(mR\sqrt{2}\right)\Bigr]\Biggr]  
+\left(mR\right)^{3}\sin\left(mR\sqrt{2}\right)\Biggl[2\sqrt{2} \ {{\textrm{Ci}}}\left(mR\sqrt{2}\right)\nonumber\\
&
&-\left(mR\right)\Bigl[\pi-2 \ {\textrm{Si}}\left(mR\sqrt{2}\right)\Bigr]\Biggr]\Biggr\} \ .
\end{eqnarray}

\begin{figure}[!h]
\centering 
\includegraphics[scale=0.45]{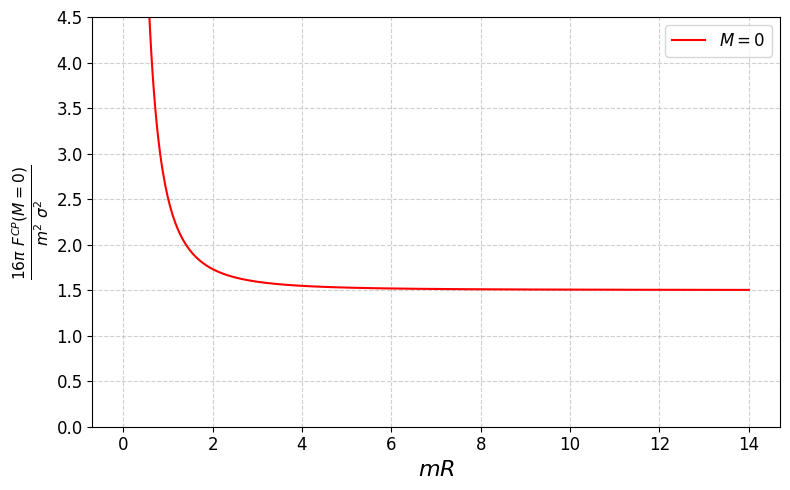} 
\caption{Interaction force $F^{CP}$, given by (\ref{FPC3}), for $M = 0$ and multiplied by the factor $\frac{16\pi}{m^2 \sigma^2}$, as a function of the dimensionless parameter $mR$. The force diverges positively as $mR \to 0$ and decreases monotonically, approaching a constant positive value for large $mR$. This indicates the repulsive nature of the interaction, consistent with Figure~\ref{grafico5}.}
\label{grafico6}
\end{figure}

The first term on the right-hand side of Eq.~(\ref{FPC3}) corresponds to the standard Coulomb interaction between the scalar charge $\sigma$ and its image, located at a distance $2R$ apart. The second term represents a non-local correction, whose behavior is described by Eq.~(\ref{delta2}). It is observed that for $mR > 0$, the correction function $\Delta_{2}(mR)$ is positive and increases with $mR$. Moreover, in the limit $mR \to 0$, the correction vanishes, i.e., $\Delta_{2}(mR) = 0$, as expected. The plot in Fig.~\ref{grafico6} illustrates the general behavior of this interaction force. 

The MIT bag model implements confinement through boundary conditions and a vacuum energy cost proportional to the bag volume~\cite{Chodos1974,Chodos1974b}, in contrast to the Cornell potential, where confinement arises from a linearly growing interaction energy. In the present nonlocal model, infrared effects generated by the rational operator lead to long-range contributions that resemble the Cornell picture, while the presence of boundaries introduces features qualitatively analogous to bag-like confinement.

Figure~\ref{grafico6} shows that the interaction force given by Eq.~(\ref{FPC3}) is consistently repulsive, since the expression within brackets on the right-hand side remains positive for all values of $mR$. In the regime where $mR$ is small, the correction term satisfies $\Delta_{2}(mR) \approx 0$, and the force in Eq.~(\ref{FPC3}) behaves similarly to the classical Coulomb interaction. Notably, as $mR$ increases, the force between the charge and the Dirichlet plane tends toward a finite limiting value, given by
\begin{equation}
\label{FPC4}   
\lim_{mR\to\infty}\frac{[F^{CP}(M=0)] \cdot 16\pi}{\sigma^{2}m^{2}}=\frac{3}{2} 
\quad \Rightarrow \quad F^{CP}(M=0)=\frac{3\sigma^{2}}{32\pi}m^{2} \ ,
\end{equation}
This asymptotic behavior contrasts with that of conventional scalar field theory, in which the interaction force vanishes as $mR$ increases. Consequently, for large values of $mR$, the force described by Eq.~(\ref{FPC3}) exhibits long-range behavior.

It is worth mentioning that the interaction force given in Eq.~(\ref{FPC3}) is equivalent (up to an overall minus sign) to the one obtained in Ref.~\cite{CL4} for the gauge field, where the interaction between a point-like charge and a perfectly conducting plate is mediated by the electromagnetic counterpart of the model described in Eq.~(\ref{model1}). This finding reveals that the correspondence between the scalar field in the presence of a Dirichlet plane and the gauge field in the presence of a perfectly conducting plate remains valid even when both theories are subject to the same type of non-local modifications. This result is quite remarkable, as in both cases we can define a dimensionless parameter $mR$, yet the functional form of the energy cannot be anticipated prior to performing the calculations.

To examine the applicability of the image method, we analyze the expression given in Eq.~(\ref{FCC1}) for the particular case in which $a = 2R$, yielding
\begin{equation}
\label{FIM}   
F^{CC}\left(M=0\right)=\frac{\sigma^{2}}{16\pi R^{2}}\left[1+2\left(mR\right)^{2}\right] \ .
\end{equation}

It is clear that Eq.~(\ref{FIM}) differs from Eq.~(\ref{FPC3}), indicating that the image method does not hold for the model described by Eq.~(\ref{model1}), in which the field is subject to the Dirichlet boundary condition given in Eq.~(\ref{Dirich}). A comparable situation arises in non-local electrodynamics in the presence of a conducting surface, where the image method is likewise not applicable~\cite{CL4}.

In the following case, we investigate the interaction between a static point-like scalar charge and the Dirichlet plane for the configuration where $M = 2m$. Substituting Eqs.~(\ref{Source}) and~(\ref{prpoplate3}) into Eq.~(\ref{Energy}), we obtain
\begin{eqnarray}
\label{ECP10}   
E^{CP}\left(M=2m\right)=\frac{\sigma^{2}}{32\pi^{2}}\int d^{2}{\bf {p}}_{\parallel}\frac{e^{-2R\sqrt{{\bf {p}}_{\parallel}^{2}+2m^{2}}}}{\left(2{\bf {p}}_{\parallel}^{2}+3m^{2}\right)\left[{\bf {p}}_{\parallel}^{2}+2m^{2}\right]^{3/2}}\left[2{\bf {p}}_{\parallel}^{2}+m^{2}\left(3-R\sqrt{{\bf {p}}_{\parallel}^{2}+2m^{2}} \ \right)\right]^{2} \ ,
\end{eqnarray}
and after suitable manipulations, we arrive at
\begin{eqnarray}
\label{ECPSM2mint}   
E^{CP}\left(M=2m\right)&=&\frac{m\sigma^{2}}{16\pi\sqrt{2}}\Biggl[\int_{1}^{\infty}dv\left(4v^{2}-1\right)
\frac{e^{-2\sqrt{2} \ (mR) v}}{v^{2}} \nonumber\\
&&-2\sqrt{2} \ (mR)\int_{1}^{\infty}dv\frac{e^{-2\sqrt{2} \ (mR) v}}{v}+2\left(mR\right)^{2}\int_{1}^{\infty}dv\frac{e^{-2\sqrt{2} \ (mR) v}}{\left(4v^{2}-1\right)}\Biggr] \ .
\end{eqnarray}

By evaluating the integrals, we obtain
\begin{eqnarray}
\label{ECPSM2m}   
E^{CP}\left(M=2m\right)=\frac{\sigma^{2}}{16\pi R} \ e^{-2\sqrt{2} \ mR}\left(1+\Delta_{3}\left(mR\right)\right) \;,
\end{eqnarray}
where
\begin{eqnarray}
\label{delta3}    
\Delta_{3}\left(mR\right)=-\frac{mR}{\sqrt{2}}+\frac{\left(mR\right)^{3}}{2\sqrt{2}}\left[e^{\sqrt{2} mR}Ei\left(1,\sqrt{2} \ mR\right)-e^{3\sqrt{2} mR}Ei\left(1,3\sqrt{2} \ mR\right)\right] \ ,
\end{eqnarray}
and $Ei$ is the exponential integral function defined as~\cite{Arfken}:
\begin{equation}
\label{Ei1}
Ei(n,s)=\int_{1}^{\infty}\frac{e^{-ts}}{t^{n}}\ dt, \quad \Re(s)>0, \quad n=0,1,2,\dots \ .
\end{equation}

The expression in Eq.~(\ref{ECPSM2m}) gives the exact interaction energy between a Dirichlet plane and a point-like scalar charge in the case $M = 2m$. The first term on the right-hand side represents the standard Yukawa interaction between the charge and its image located at a distance $2R$, while the second term, given by $\Delta_{3}(mR)$, encodes the non-local corrections governed by the parameter $m$.

\begin{figure}[htbp]
    \centering
    \begin{subfigure}[b]{0.40\textwidth}
        \includegraphics[width=\textwidth]{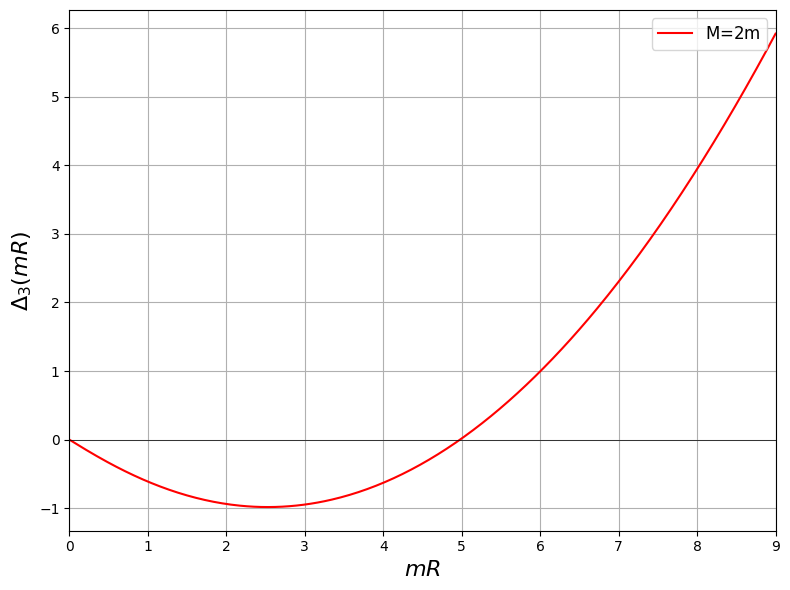}
        \caption{Plot of $\Delta_{3}\left(mR\right)$ from Eq.~(\ref{delta3}).}
        \label{fig:sub71}
    \end{subfigure}
    \hfill
    \begin{subfigure}[b]{0.48\textwidth}
        \includegraphics[width=\textwidth]{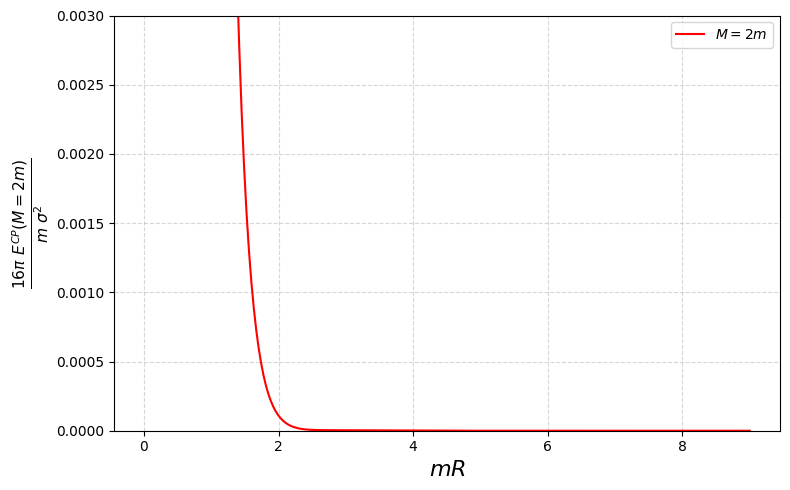}
        \caption{Normalized energy $E^{CP}(M=2m)$ multiplied by $\frac{16\pi}{m\sigma^{2}}$ as a function of $mR$.}
        \label{fig:sub72}
    \end{subfigure}
    \caption{Behavior of the interaction energy given in Eq.~(\ref{ECPSM2m}) for the case \( M = 2m \). (a) Function \( \Delta_3(mR) \), indicating the non-local contribution. (b) Normalized energy, highlighting the transition between Yukawa-like and long-range behavior.}
    \label{grafico7}
\end{figure}

To better understand the behavior of the interaction energy in Eq.~(\ref{ECPSM2m}), Fig.~\ref{grafico7} displays its dependence on $mR$. More precisely, Fig.~\ref{fig:sub71} shows the function $\Delta_{3}(mR)$, and Fig.~\ref{fig:sub72} shows the energy multiplied by $\frac{16\pi}{m\sigma^2}$.

From Fig.~\ref{grafico7}, we observe that $E^{CP}(M=2m)$, given by (\ref{ECPSM2m}), remains positive except at the point where the expression vanishes. In the interval $0 < mR \lesssim 2.6$, the function $\Delta_{3}(mR)$ is negative, and its magnitude increases with $mR$. At $mR \approx 2.6$, we find $\Delta_{3}(mR) \approx -1$, making the energy vanish due to cancellation between the Yukawa and non-local contributions. Between $2.6 \lesssim mR \lesssim 5$, the function $\Delta_{3}(mR)$ remains negative but begins to increase, reaching zero around $mR \approx 5$. Beyond this point, $\Delta_{3}(mR)$ becomes positive and grows, meaning the non-local correction dominates for $mR \gg 5$, characterizing a long-range interaction decaying with distance, as shown in Fig.~\ref{fig:sub72}.

Furthermore, the interaction force obtained from (\ref{ECPSM2m}) reads
\begin{eqnarray}
\label{FCPSM2m}    
F^{CP}\left(M=2m\right)=\frac{\sigma^{2}}{16\pi R^{2}}\left(1 + 2 \sqrt{2} \, m R\right)e^{-2\sqrt{2} \ mR}\Bigl(1-\Delta_{4}\left(mR\right)\Bigr) \ ,
\end{eqnarray}
with, 
\begin{eqnarray}
\label{delta4}
\Delta_{4}(mR)&=&\frac{2 (m R)^2}{1 + 2 \sqrt{2} \, m R} + 
\frac{(m R)^3}{2 \sqrt{2} \left(1 + 2 \sqrt{2} \, m R\right)} 
\Biggl[
\left(2 - \sqrt{2} \, m R\right) e^{\sqrt{2} \, m R} \, {Ei}\left(1,\sqrt{2} \, m R\right)\nonumber\\
&
&- \left(2 + \sqrt{2} \, m R\right) e^{3 \sqrt{2} \, m R} \, {Ei}\left(1,3 \sqrt{2} \, m R\right)
\Biggr] \ .
\end{eqnarray}

From Eq.~(\ref{FCPSM2m}), the first term in parentheses on the right-hand side represents the standard Yukawa interaction between the scalar charge $\sigma$ and its image, located at a distance $2R$. The second term corresponds to a non-local contribution, whose behavior is governed by the function $\Delta_{4}(mR)$ defined in Eq.~(\ref{delta4}).

\begin{figure}[htbp]
    \centering
    \begin{subfigure}[b]{0.45\textwidth}
        \includegraphics[width=\textwidth]{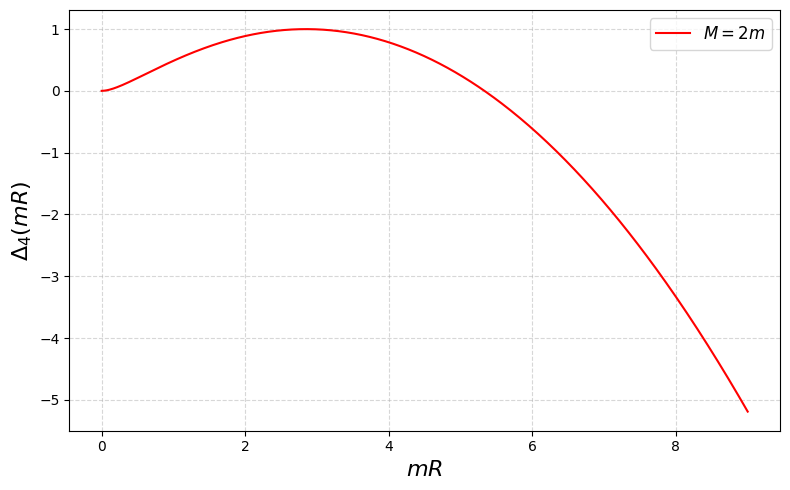}
        \caption{Plot for $\Delta_{4}\left(mR\right)$ in Eq. (\ref{delta4}).}
        \label{fig:sub81}
    \end{subfigure}
    \hfill
    \begin{subfigure}[b]{0.45\textwidth}
        \includegraphics[width=\textwidth]{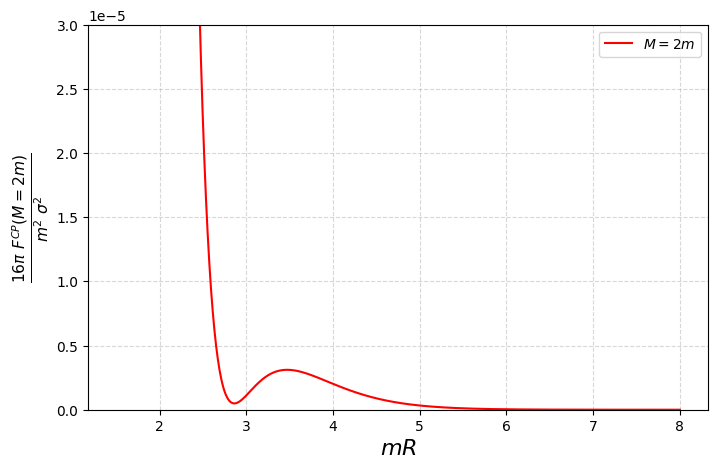}
        \caption{Force (\ref{FCPSM2m}) multiplied by $\frac{16\pi}{m^{2}\sigma^{2}}$ 
as a function of $mR$.}
        \label{fig:sub82}
    \end{subfigure}
    \caption{Behavior of the interaction force (\ref{FCPSM2m}) for the case \( M = 2m \). 
    (a) defined in Eq.~(\ref{delta4}), which determines the distance-dependent correction to the force. 
    (b) Normalized force \( \frac{16\pi}{m^2\sigma^2} F^{CP}(M=2m) \) as a function of \( mR \), exhibiting a sharp peak and a local minimum at short distances, followed by a rapid decay. 
    The repulsive nature of the force is consistent with the energy profile in Fig.~\ref{grafico7}.}.
    \label{grafico8}
\end{figure}

To further investigate the characteristics of the interaction force described by Eq.~(\ref{FCPSM2m}), Fig.~\ref{grafico8} provides a graphical representation of its behavior. Specifically, Fig.~\ref{fig:sub81} shows the function $\Delta_{4}(mR)$, while Fig.~\ref{fig:sub82} presents the interaction force given by Eq.~(\ref{FCPSM2m}), scaled by the factor $\frac{16\pi}{m^{2}\sigma^{2}}$, as a function of the dimensionless parameter $mR$.

As shown in Fig.~\ref{grafico8}, the normalized force (\ref{FCPSM2m}) is repulsive, since the term enclosed in parentheses on the right-hand side remains positive for all values of $mR>0$. Moreover, it exhibits a nontrivial distance dependence, with a pronounced local minimum at $mR\simeq2.85$, where the normalized force takes the very small value $\simeq1.7\times10^{-4}$. As the separation increases, the normalized force grows and reaches a local maximum at $mR\simeq3.50$, with value $\simeq2.5\times10^{-1}$. This behavior indicates a strong suppression of the normalized force around $mR\simeq2.85$, followed by a rapid increase up to a peak near $mR\simeq3.50$. Beyond this point, the normalized force gradually decreases, approaching zero asymptotically at large separations.

In order to assess the validity of the image method, we consider Eq.~(\ref{FCCS2}) for the special case where $\sigma_{1} = -\sigma_{2} = \sigma$ and $a = 2R$,
\begin{eqnarray}
\label{FCCS2R}    
F^{CC}\left(M=2m\right)=\frac{\sigma^{2}}{16\pi R^{2}}\left(1+2{\sqrt{2}} \ mR\right)e^{-2\sqrt{2} \ mR}\left[1-\frac{2\left(mR\right)^{2}}{\left(1+{2\sqrt{2}} \ mR\right)}\right] \ .
\end{eqnarray}
We observe that Eq.~(\ref{FCCS2R}) differs from Eq.~(\ref{FCPSM2m}), indicating that the image method is not valid for the model described by Eq.~(\ref{model1}) when subject to the Dirichlet boundary condition given in Eq.~(\ref{Dirich}), in the configuration where $M = 2m$.


In the third scenario, we analyze the configuration in which $0 < \frac{4m^{2}}{M^{2}} < 1$. Under this condition, two distinct modes arise, characterized by different non-zero effective masses denoted by $m_{+}$ and $m_{-}$, which are defined in Eqs.~(\ref{poleS}) and~(\ref{poless}), respectively.

Substituting (\ref{Source}) and (\ref{proplate4}) in (\ref{Energy}), we obtain
\begin{eqnarray}
\label{ECPSm+m-1}     
E^{CP}\left(m_{+},m_{-}\right)&=&\frac{\sigma^{2}}{16\pi^{2}}\frac{1}{\left(m_{+}^{2}-m_{-}^{2}\right)}\int d^{2}{\bf{p}}_{\parallel}\frac{\sqrt{{\bf{p}}_{\parallel}^{2}+m_{+}^{2}}\sqrt{{\bf{p}}_{\parallel}^{2}+m_{-}^{2}}}{\left[\left(m_{+}^{2}-m^{2}\right)\sqrt{{\bf{p}}_{\parallel}^{2}+m_{-}^{2}}-\left(m_{-}^{2}-m^{2}\right)\sqrt{{\bf{p}}_{\parallel}^{2}+m_{+}^{2}} \ \right]}\nonumber\\
&
&\times\Biggl[\left(m_{+}^{2}-m^{2}\right)\frac{e^{-R\sqrt{{\bf{p}}_{\parallel}^{2}+m_{+}^{2}}}}{\sqrt{{\bf{p}}_{\parallel}^{2}+m_{+}^{2}}}-\left(m_{-}^{2}-m^{2}\right)\frac{e^{-R\sqrt{{\bf{p}}_{\parallel}^{2}+m_{-}^{2}}}}{\sqrt{{\bf{p}}_{\parallel}^{2}+m_{-}^{2}}} \ \Biggr]^{2} \ .
\end{eqnarray}

By using polar coordinates, integrating over the solid angle, performing the change of the integration variable
$u=\frac{\mid{\bf{p}}_{\parallel}\mid}{m}$, and using the definitions (\ref{m+m-li}), the interaction energy reads
\begin{eqnarray}
\label{ECPSm+m-2}
E^{CP}\left(m'_{+},m'_{-}\right)&=&\frac{\sigma^{2}}{8\pi}\frac{m}{\left[(m'_{+})^{2}-(m'_{-})^{2}\right]}\int_{0}^{\infty} du \frac{u\sqrt{u^{2}+{(m'_{+})^{2}}}\sqrt{u^{2}+{(m'_{-})^{2}}}}{\left\{\left[(m'_{+})^{2}-1\right]\sqrt{u^{2}+{(m'_{-})^{2}}}-\left[(m'_{-})^{2}-1\right]\sqrt{u^{2}+{(m'_{+})^{2}}} \ \right\}}\nonumber\\
&
&\times\Biggl[\left[(m'_{+})^{2}-1\right]^{2} \ \frac{e^{-2mR\sqrt{u^{2}+{(m'_{+})^{2}}}}}{u^{2}+{(m'_{+})^{2}}}\nonumber\\
&
&-2\left[(m'_{+})^{2}-1\right]\left[(m'_{-})^{2}-1\right]\frac{e^{-mR\left(\sqrt{u^{2}+(m'_{+})^{2}}+\sqrt{u^{2}+(m'_{-})^{2}}\right)}}{\sqrt{u^{2}+{(m'_{+})^{2}}}\sqrt{u^{2}+{(m'_{-})^{2}}}}\nonumber\\
&
&+\left[(m'_{-})^{2}-1\right]^{2} \ \frac{e^{-2mR\sqrt{u^{2}+{(m'_{-})^{2}}}}}{u^{2}+{(m'_{-})^{2}}}\Biggr] \ .
\end{eqnarray}

The force corresponding to (\ref{ECPSm+m-2}) is given by
\begin{eqnarray}
\label{FCPSm+m-}    
F^{CP}\left(m'_{+},m'_{-}\right)&=&\frac{\sigma^{2}}{4\pi}\frac{m^{2}}{\left[(m'_{+})^{2}-(m'_{-})^{2}\right]}\int_{0}^{\infty} du \frac{u\sqrt{u^{2}+{(m'_{+})^{2}}}\sqrt{u^{2}+{(m'_{-})^{2}}}}{\left\{\left[(m'_{+})^{2}-1\right]\sqrt{u^{2}+{(m'_{-})^{2}}}-\left[(m'_{-})^{2}-1\right]\sqrt{u^{2}+{(m'_{+})^{2}}} \ \right\}}\nonumber\\
&
&\times\left[\left[(m'_{+})^{2}-1\right]\frac{e^{-mR\sqrt{u^{2}+{(m'_{+})^{2}}}}}{\sqrt{u^{2}+{(m'_{+})^{2}}}}-\left[(m'_{-})^{2}-1\right]\frac{e^{-mR\sqrt{u^{2}+{(m'_{-})^{2}}}}}{\sqrt{u^{2}+{(m'_{-})^{2}}}}\right]\nonumber\\
&
&\times\left[\left[(m'_{+})^{2}-1\right]e^{-mR\sqrt{u^{2}+{(m'_{+})^{2}}}}-\left[(m'_{-})^{2}-1\right]e^{-mR\sqrt{u^{2}+{(m'_{-})^{2}}}}\right] \ .
\end{eqnarray}
We observe that both the interaction energy and the interaction force, given in Eqs.~(\ref{ECPSm+m-2}) and~(\ref{FCPSm+m-}), respectively, are exact expressions written in terms of integrals that cannot be evaluated in closed analytical form. Therefore, in order to investigate the general behavior of the plane–charge interaction, we resort to numerical methods to analyze these quantities.

Figure~(\ref{FIG55}) displays the interaction energy given in Eq.~(\ref{ECPSm+m-2}), multiplied by the factor $\frac{8\pi}{m\sigma^{2}}$, as a function of the dimensionless parameter $mR$. The plot includes four distinct scenarios corresponding to different values of the mass parameter $M$: the blue curve denotes $M = \sqrt{5}\,m$, the orange curve corresponds to $M = \sqrt{15}\,m$, the green curve represents $M = \sqrt{50}\,m$, and the red curve illustrates the case $M = \sqrt{100}\,m$.

\begin{figure}[H]
 
  \centering
  
   \begin{subfigure}[t]{0.48\textwidth}
    \centering
    \includegraphics[width=\linewidth]{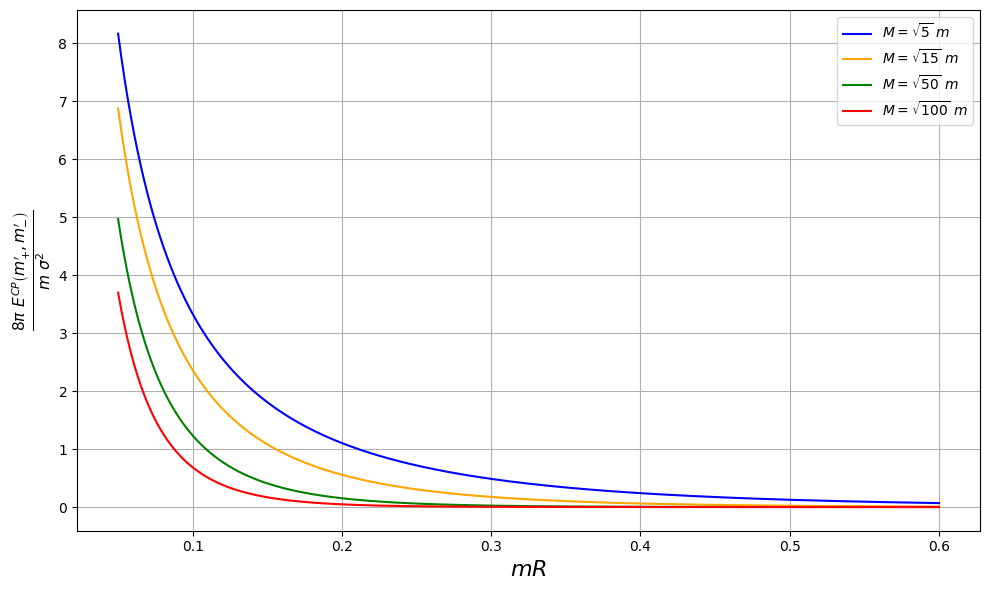}
    \caption{Small values of $mR$.}
    \label{fig:55}
  \end{subfigure}
  \hfill
  \begin{subfigure}[t]{0.48\textwidth}
    \centering
    \includegraphics[width=\linewidth]{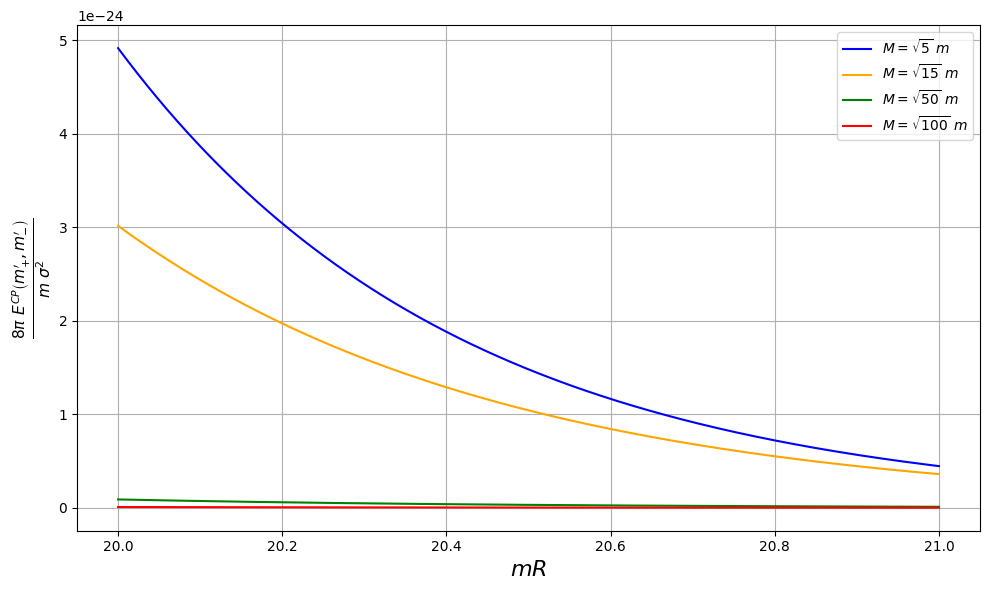}
    \caption{Interval: $20\leq mR \leq 21$.}
    \label{fig:66}
  \end{subfigure}
\hfill
\begin{subfigure}[t]{0.48\textwidth}
    \centering
    \includegraphics[width=\linewidth]{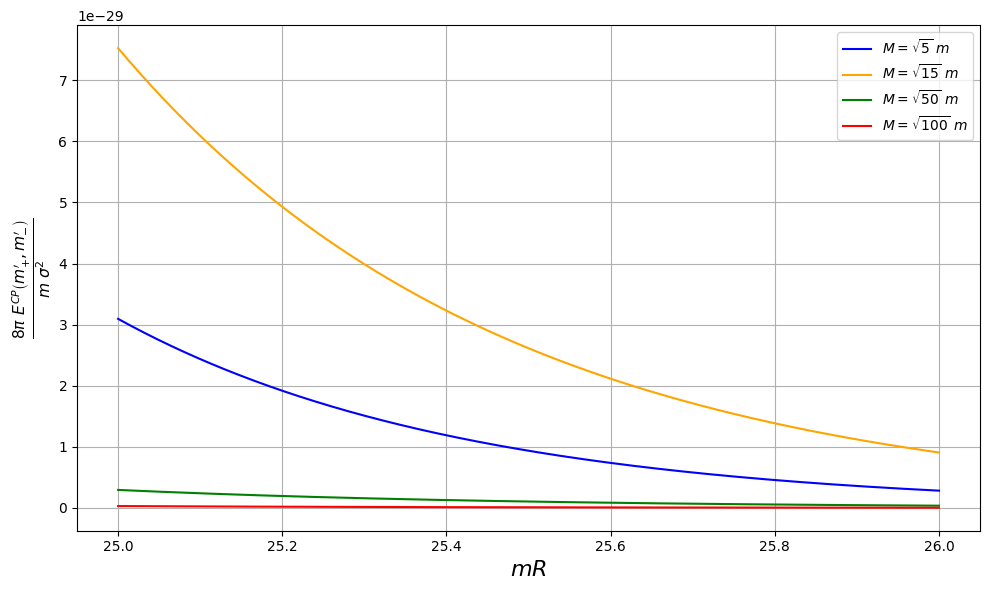}
    \caption{Interval: $25\leq mR \leq 26$.}
    \label{fig:77}
  \end{subfigure}
\hfill
\begin{subfigure}[t]{0.48\textwidth}
    \centering
    \includegraphics[width=\linewidth]{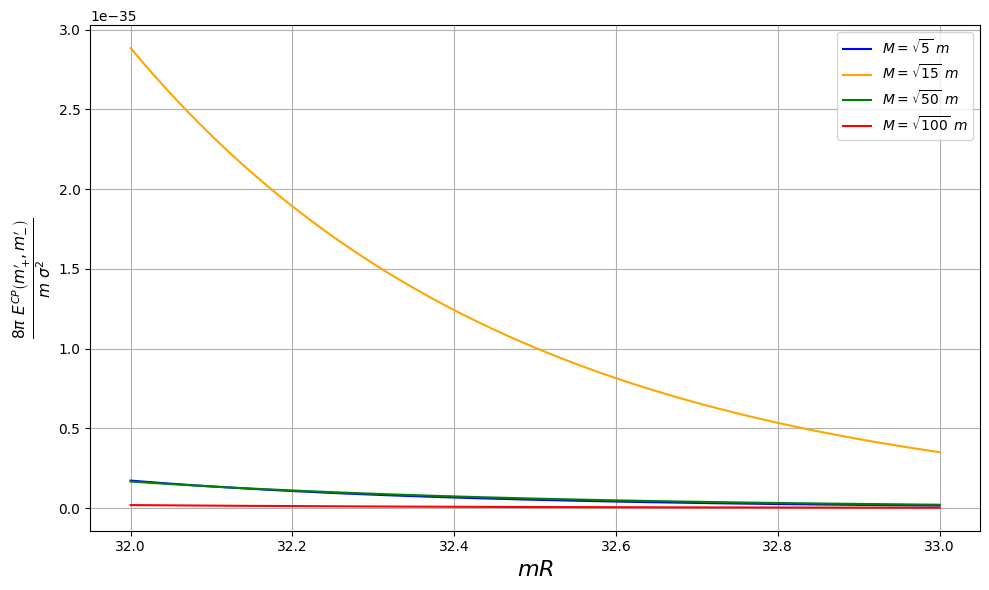}
    \caption{Interval: $32\leq mR \leq 33$.}
    \label{fig:88}
  \end{subfigure}
\hfill
\begin{subfigure}[t]{0.48\textwidth}
    \centering
    \includegraphics[width=\linewidth]{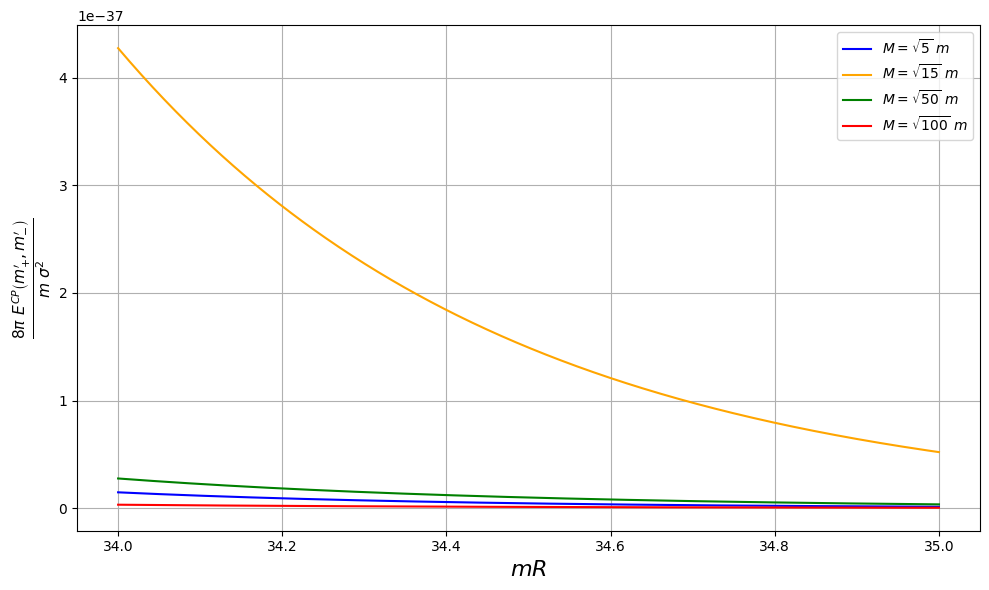}
    \caption{interval: $34\leq mR \leq 35$.}
    \label{fig:99}
  \end{subfigure}
\hfill
\begin{subfigure}[t]{0.48\textwidth}
    \centering
    \includegraphics[width=\linewidth]{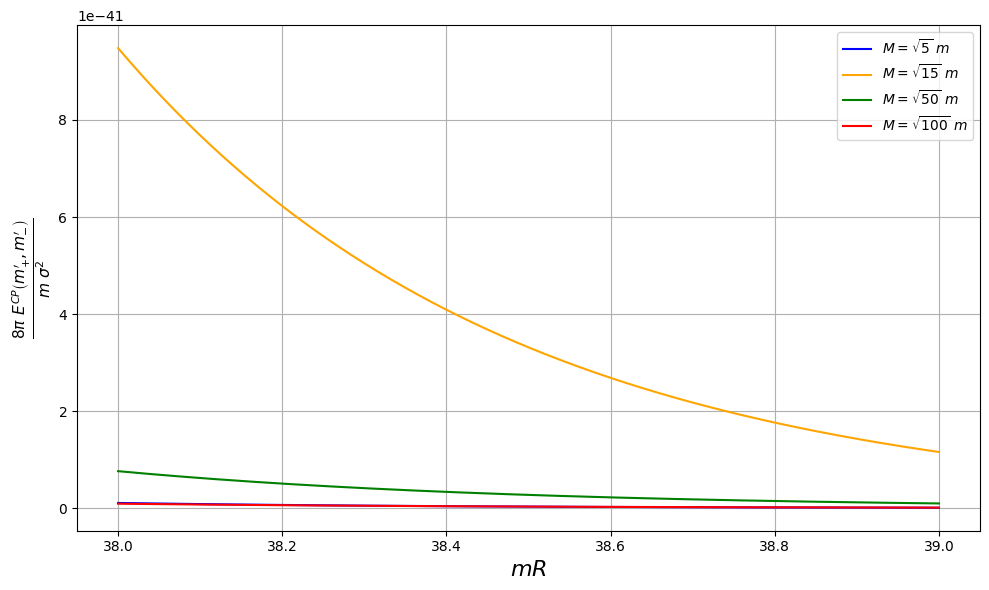}
    \caption{Interval: $38\leq mR \leq 39$.}
    \label{fig:1010}
  \end{subfigure}
\end{figure}

\begin{figure}[H]
\centering
\ContinuedFloat
    \centering
    \begin{subfigure}[t]{0.48\textwidth}
    \includegraphics[width=\linewidth]{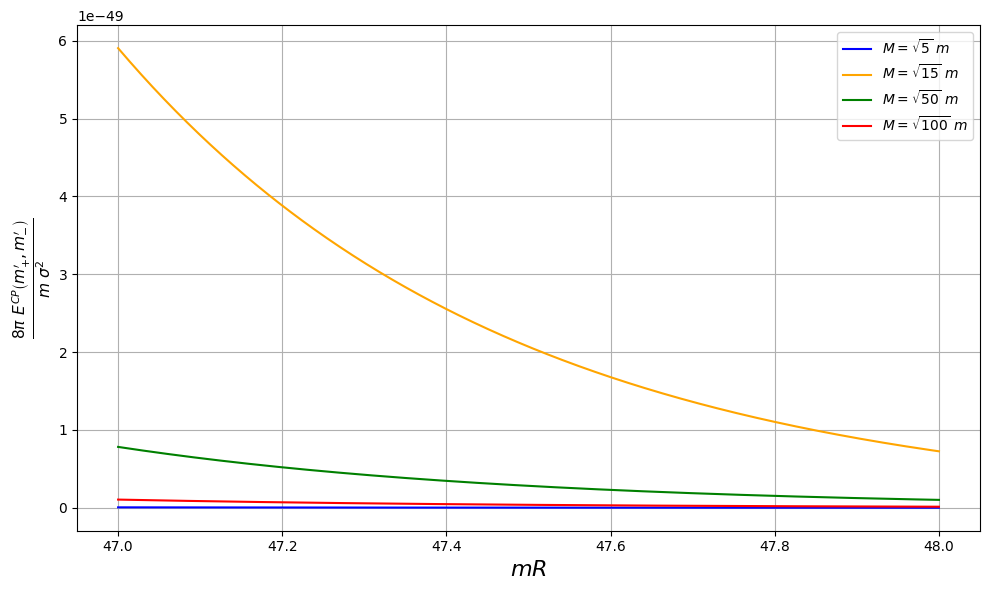}
    \caption{Interval: $47\leq mR \leq 48$.}
    \label{fig:1111}
  \end{subfigure}
\hfill
\begin{subfigure}[t]{0.48\textwidth}
    \centering
    \includegraphics[width=\linewidth]{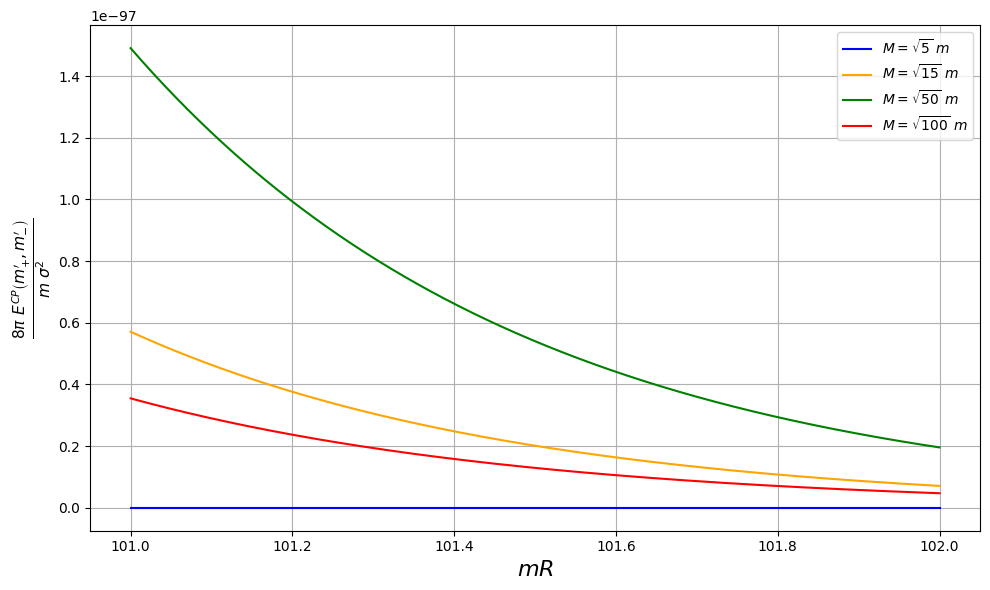}
    \caption{Interval: $101\leq mR \leq 102$.}
    \label{fig:1212}
  \end{subfigure}
\hfill
\begin{subfigure}[t]{0.48\textwidth}
    \centering
    \includegraphics[width=\linewidth]{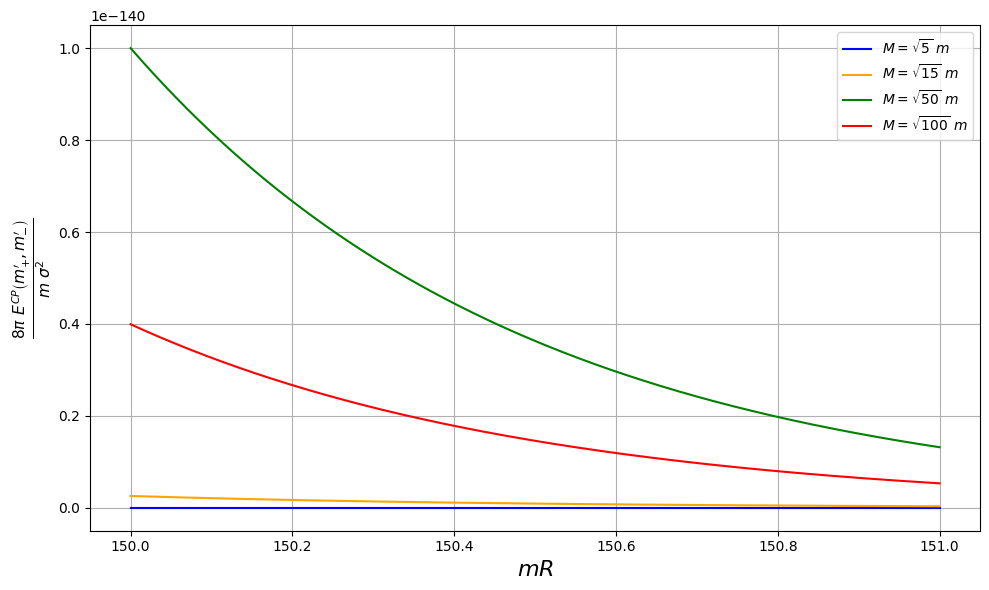}
    \caption{Interval: $150\leq mR \leq 151$.}
    \label{fig:1313}
  \end{subfigure}
\hfill
\begin{subfigure}[t]{0.48\textwidth}
    \centering
    \includegraphics[width=\linewidth]{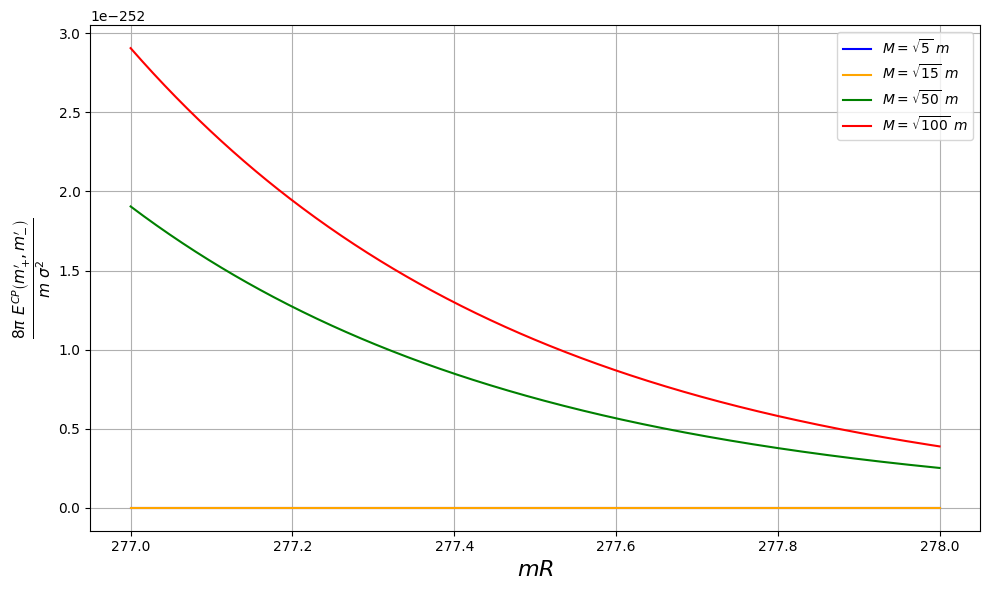}
    \caption{Interval: $277\leq mR \leq 278$.}
    \label{fig:1414}
\end{subfigure}
\hfill
\begin{subfigure}[t]{0.48\textwidth}
    \centering
    \includegraphics[width=\linewidth]{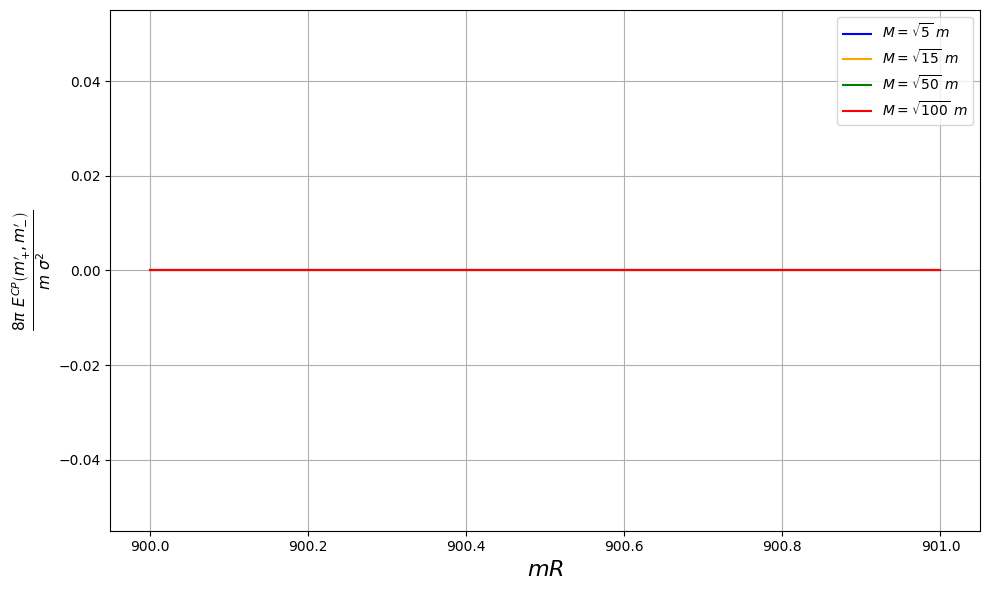}
    \caption{Interval: $900\leq mR \leq 901$.}
    \label{fig:1515}
\end{subfigure}
 
\caption{Energy described by Eq.~(\ref{ECPSm+m-2}), multiplied by \( \frac{8\pi}{m\sigma^2} \), as a function of the dimensionless distance \( mR \), for several mass ratios \( M/m \). Each figure shows the behavior in different \( mR \)-intervals, highlighting the transition from short-range divergence to long-range saturation. 
A change in the hierarchy of the curves is observed as \( mR \) increases: while lower \( M \) values dominate at short distances, the energy becomes larger for higher \( M \) in the asymptotic regime.}
 \label{FIG55}
\end{figure}

Based on Fig.~(\ref{fig:55}), we can describe the general behavior of the interaction energy~(\ref{ECPSm+m-2}), which remains positive throughout. For small fixed values of $mR$, the energy~(\ref{ECPSm+m-2}) decreases as the mass $M$ increases, as illustrated in Fig.~(\ref{fig:55}). However, as shown in Figs.~(\ref{fig:66})--(\ref{fig:1313}), this trend gradually reverses with increasing $mR$ due to the effects introduced by nonlocality. Specifically, for large values of $mR$, where nonlocal contributions dominate, the interaction energy~(\ref{ECPSm+m-2}) increases with increasing $M$ for fixed $mR$, as indicated in Fig.~(\ref{fig:1414}). All curves approach zero asymptotically for large values of $mR$, as shown in Fig.~(\ref{fig:1515}).

Figure~(\ref{FIG66}) illustrates the behavior of the interaction force~(\ref{FCPSm+m-}), multiplied by the factor $\frac{4\pi}{m^{2}\sigma^{2}}$, plotted as a function of $mR$ over different regimes, the same ones previously considered for the interaction energy~(\ref{ECPSm+m-2}).
\begin{figure}[H]
  \centering
    \begin{subfigure}[t]{0.48\textwidth}
    \centering
    \includegraphics[width=\linewidth]{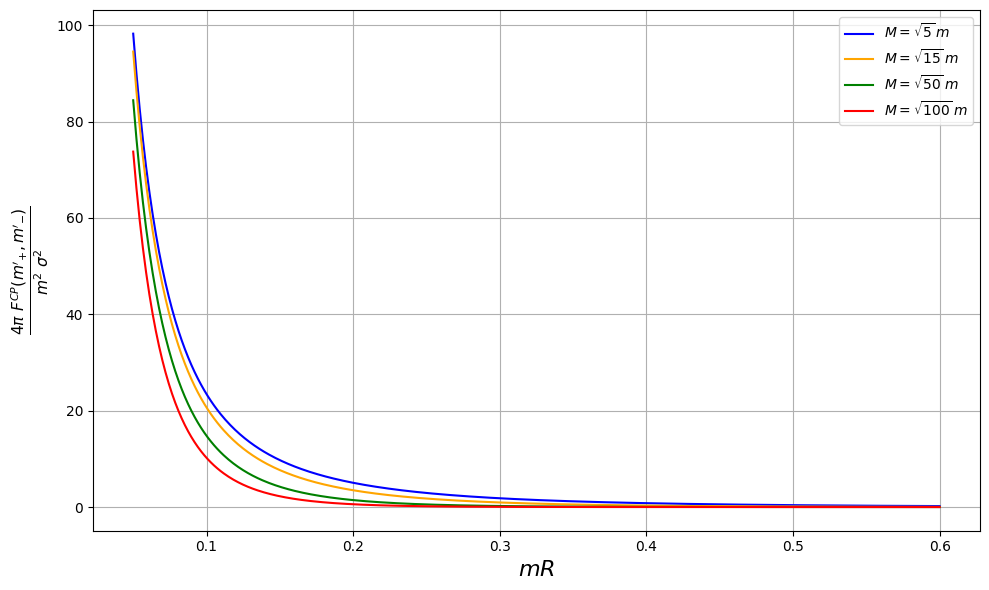}
    \caption{Small values of $mR$.}
    \label{fig:555}
  \end{subfigure}
  \hfill
  \begin{subfigure}[t]{0.48\textwidth}
    \centering
    \includegraphics[width=\linewidth]{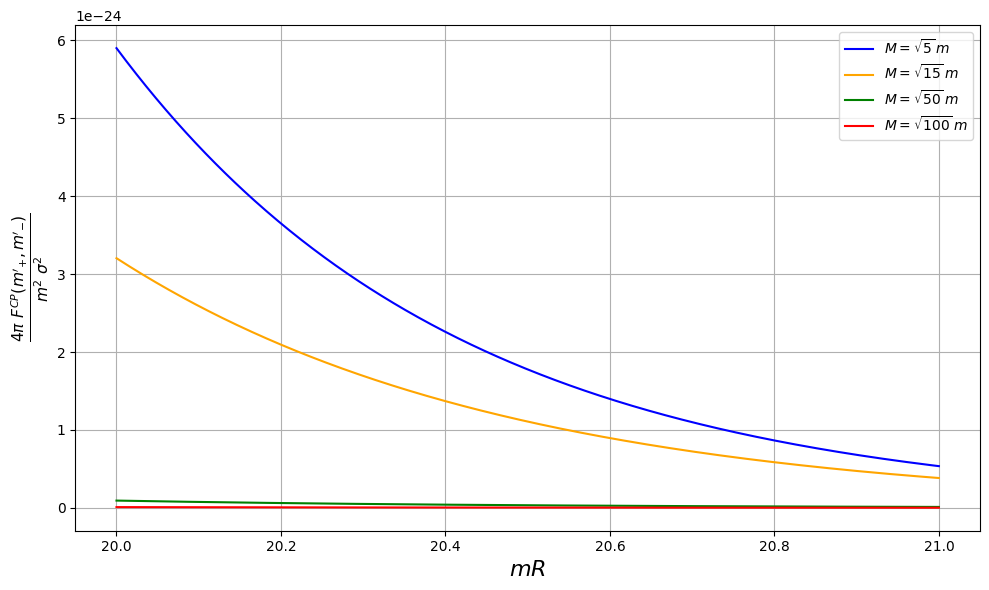}
    \caption{Interval: $20\leq mR \leq 21$.}
    \label{fig:666}
  \end{subfigure}
\hfill
\begin{subfigure}[t]{0.48\textwidth}
    \centering
    \includegraphics[width=\linewidth]{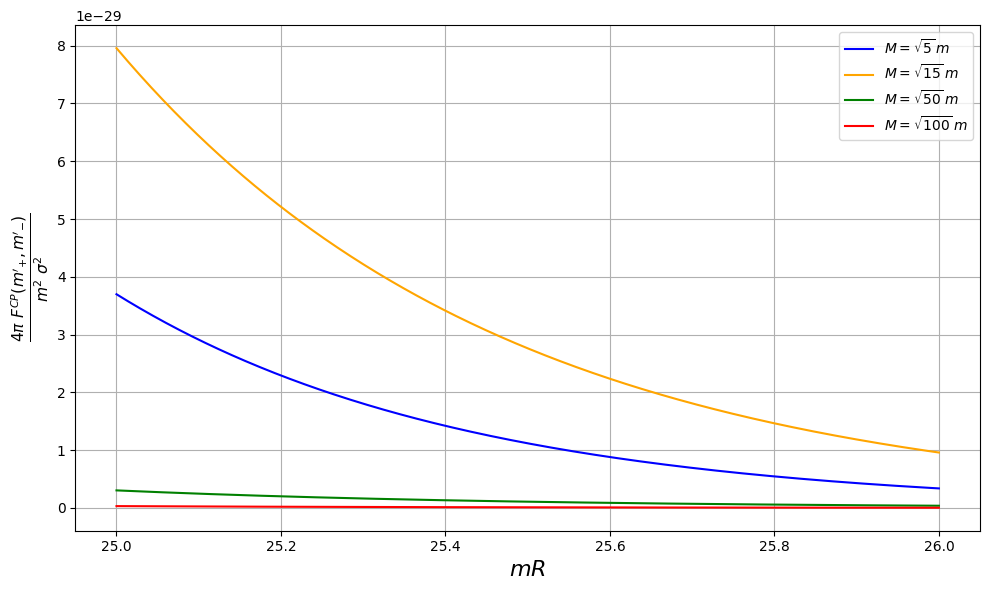}
    \caption{Interval: $25\leq mR \leq 26$.}
    \label{fig:777}
  \end{subfigure}
\hfill
\begin{subfigure}[t]{0.48\textwidth}
    \centering
    \includegraphics[width=\linewidth]{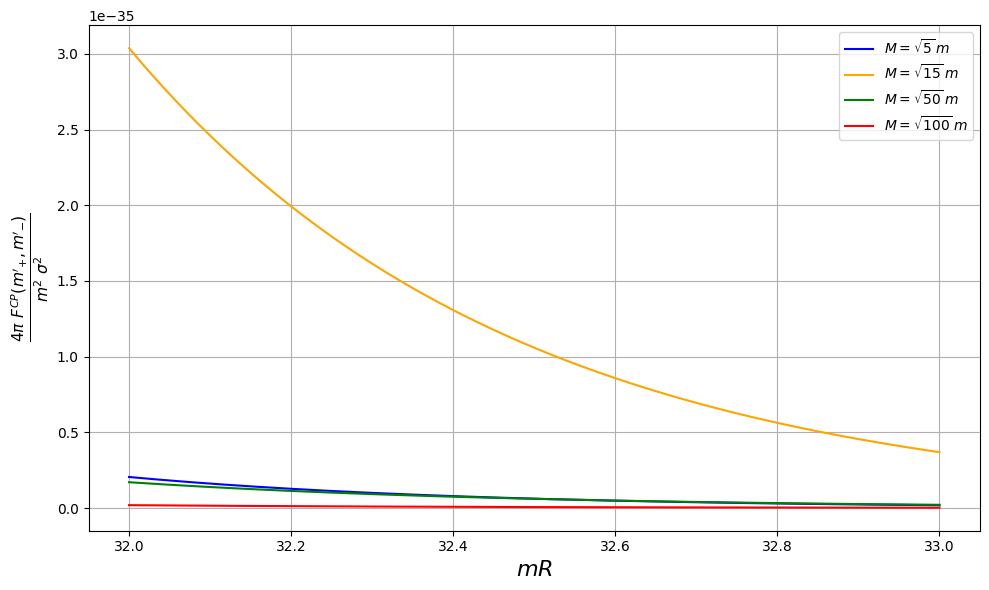}
    \caption{Interval: $32\leq mR \leq 33$.}
    \label{fig:888}
  \end{subfigure}
\hfill
\begin{subfigure}[t]{0.48\textwidth}
    \centering
    \includegraphics[width=\linewidth]{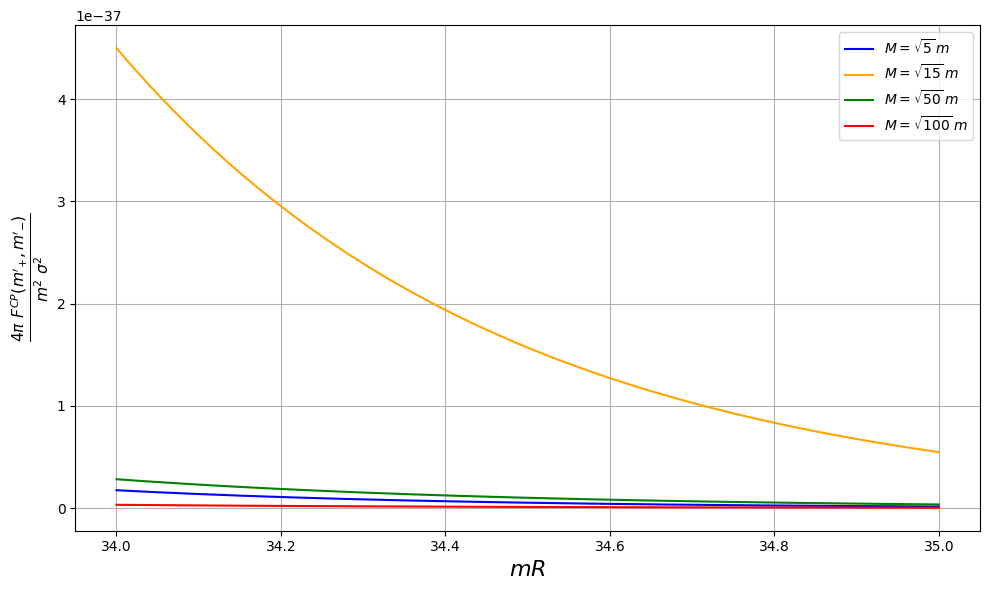}
    \caption{interval: $34\leq mR \leq 35$.}
    \label{fig:999}
  \end{subfigure}
\hfill
\begin{subfigure}[t]{0.48\textwidth}
    \centering
    \includegraphics[width=\linewidth]{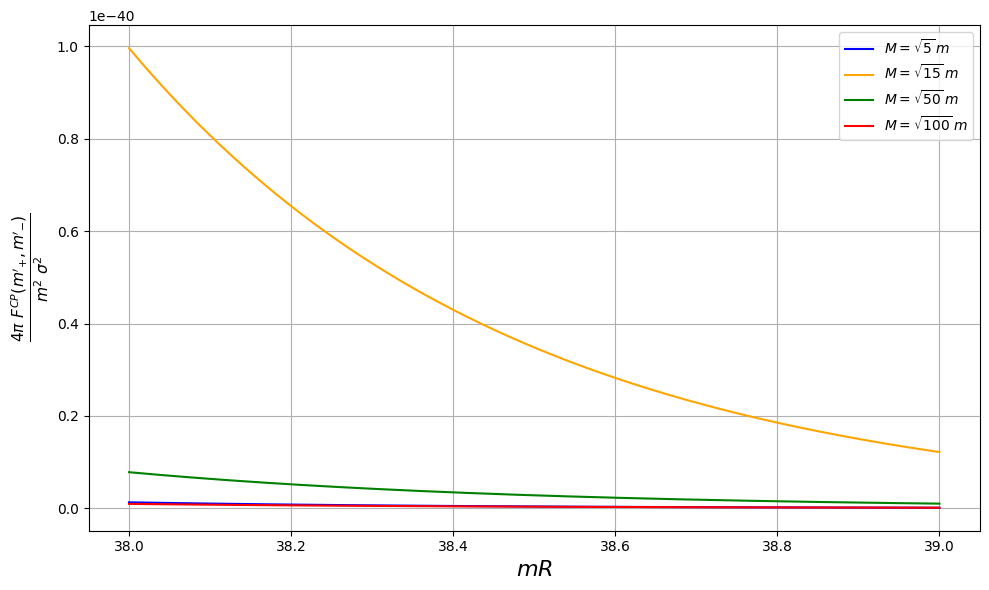}
    \caption{Interval: $38\leq mR \leq 39$.}
    \label{fig:101010}
  \end{subfigure}
\end{figure}

\begin{figure}[H]
\centering
\ContinuedFloat
    \centering
    \begin{subfigure}[t]{0.48\textwidth}
    \includegraphics[width=\linewidth]{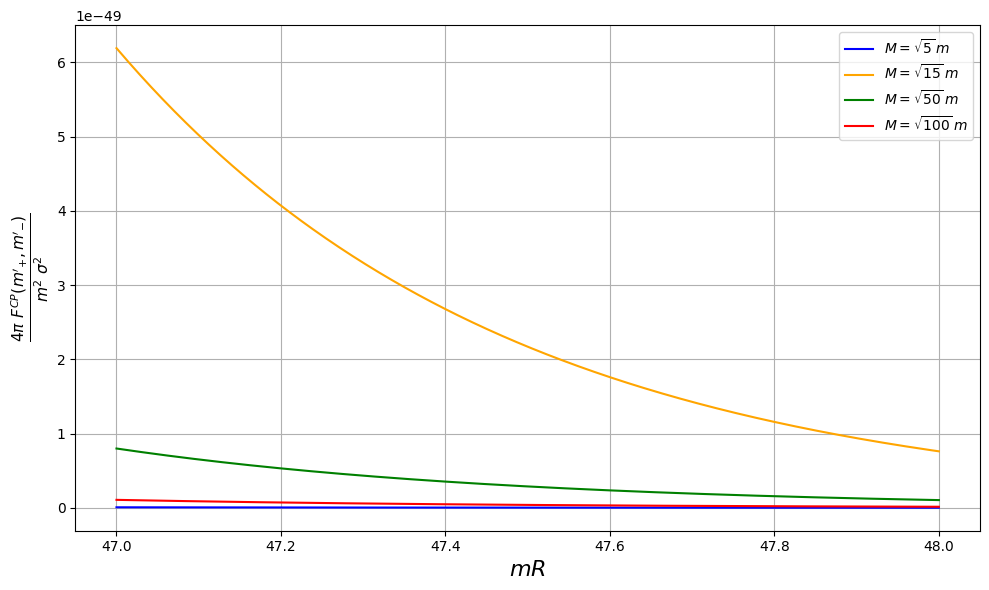}
    \caption{Interval: $47\leq mR \leq 48$.}
    \label{fig:111111}
  \end{subfigure}
\hfill
\begin{subfigure}[t]{0.48\textwidth}
    \centering
    \includegraphics[width=\linewidth]{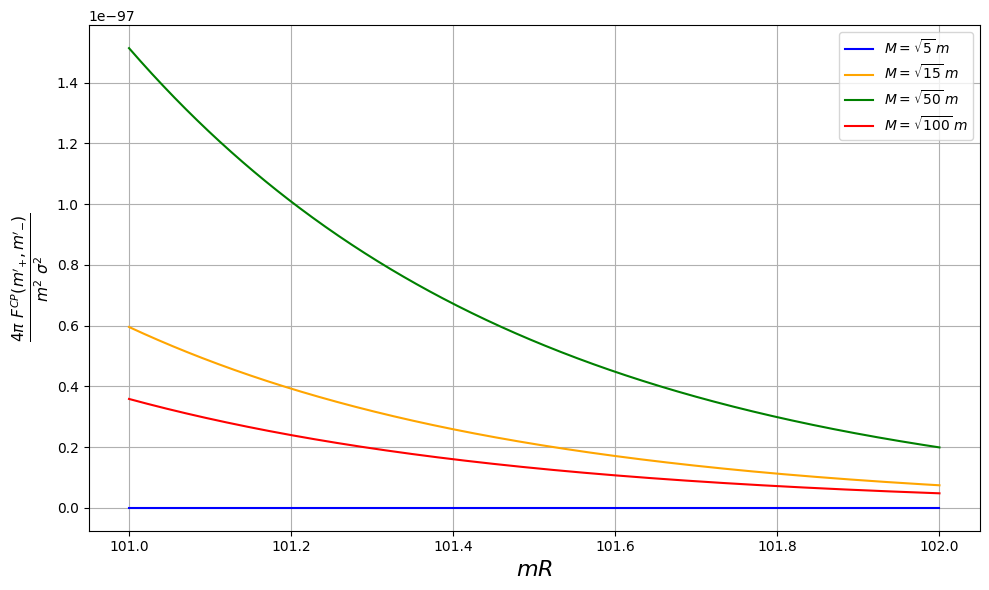}
    \caption{Interval: $101\leq mR \leq 102$.}
    \label{fig:121212}
  \end{subfigure}
\hfill
\begin{subfigure}[t]{0.48\textwidth}
    \centering
    \includegraphics[width=\linewidth]{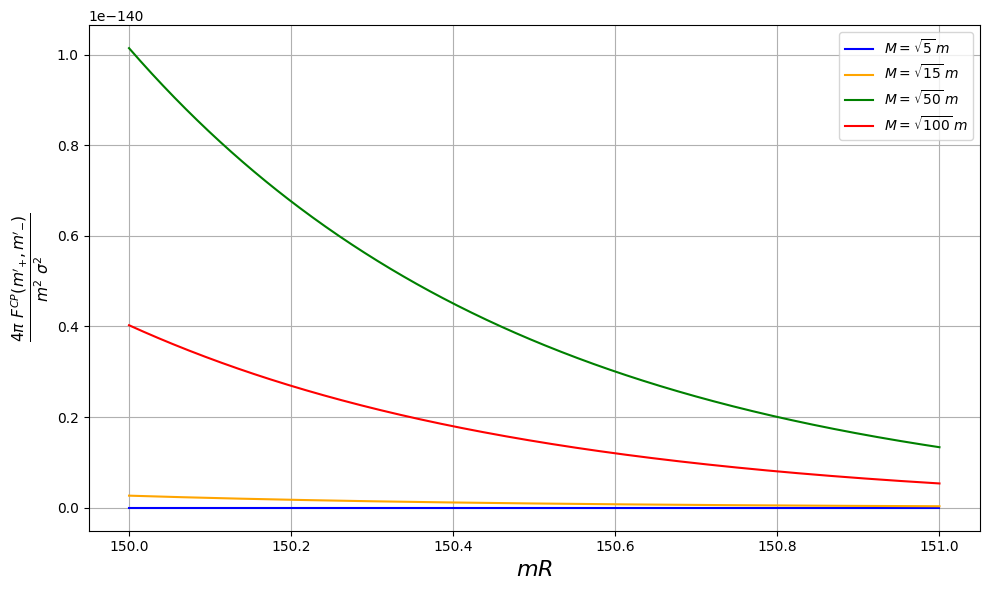}
    \caption{Interval: $150\leq mR \leq 151$.}
    \label{fig:131313}
  \end{subfigure}
\hfill
\begin{subfigure}[t]{0.48\textwidth}
    \centering
    \includegraphics[width=\linewidth]{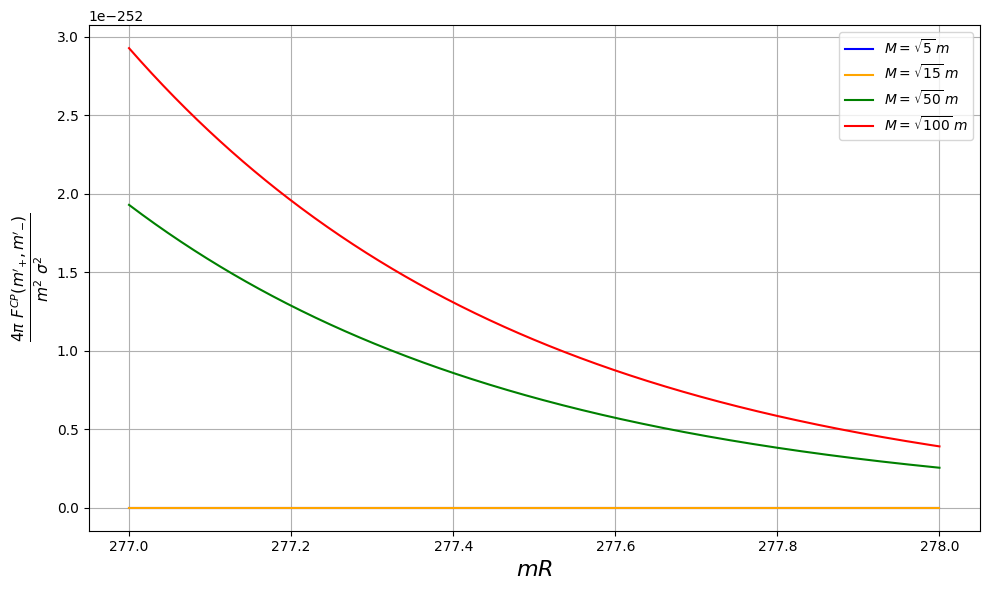}
    \caption{Interval: $277\leq mR \leq 278$.}
    \label{fig:141414}
\end{subfigure}
\hfill
\begin{subfigure}[t]{0.48\textwidth}
    \centering
    \includegraphics[width=\linewidth]{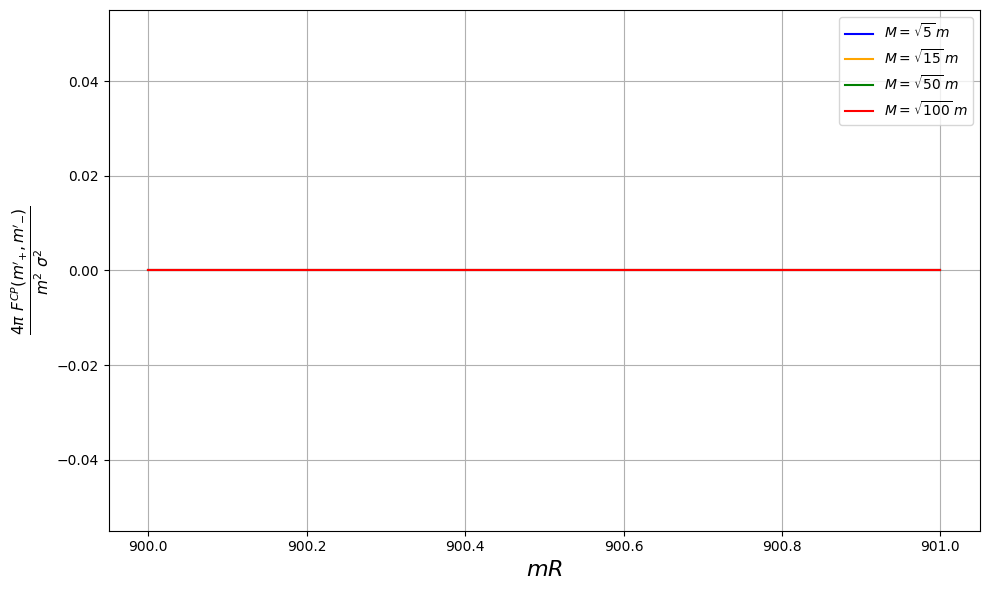}
    \caption{Interval: $900\leq mR \leq 901$.}
    \label{fig:151515}
\end{subfigure}
\caption{Force described by Eq.~(\ref{FCPSm+m-}), multiplied by \( \frac{4\pi}{m^2\sigma^2} \), as a function of the dimensionless distance \( mR \), for various mass ratios \( M/m \). The figures display different \( mR \)-intervals and illustrate the rapid decay of the repulsive interaction force with distance. For small \( mR \), lower $M$ mass values yield stronger forces, while for large \( mR \), the dominance shifts to higher $M/m$ mass ratios. The force vanishes asymptotically, confirming the short-range nature of the interaction.}
\label{FIG66}
\end{figure}

From the analysis of Fig.~(\ref{FIG66}), we can describe the overall behavior of the interaction force~(\ref{FCPSm+m-}), which is always repulsive for all values of $mR$. From Fig.~(\ref{fig:555}), we observe that for small fixed values of $mR$, larger values of the mass $M$ lead to smaller interaction forces. However, as shown in Figs.~(\ref{fig:666})--(\ref{fig:131313}), this behavior gradually reverses as $mR$ increases, due to the effects introduced by nonlocality through the parameter $m$. Consequently, for large fixed values of $mR$, where nonlocal contributions become dominant, the force~(\ref{FCPSm+m-}) increases with increasing $M$, as indicated in Fig.~(\ref{fig:141414}). From this point onward, this trend persists, and the plane–charge interaction force approaches zero asymptotically as $mR$ increases, as shown in Fig.~(\ref{fig:151515}).

As a final remark, we point out that it is straightforward to verify that the interaction force given by Eq.~(\ref{FCCS6}), in the case where $\sigma_{1} = -\sigma_{2} = \sigma$ and $a = 2R$, exhibits a behavior distinct from the charge-plate interaction force in Eq.~(\ref{FCPSm+m-}). Therefore, the image method is not applicable in the situation where $0 < \frac{4m^{2}}{M^{2}} < 1$, under Dirichlet boundary condition as specified in Eq.~(\ref{Dirich}).

\section{Conclusions}
\label{V}

In this work, we have investigated classical non-local effects in scalar field theory arising from the presence of external sources and a material boundary. By introducing a non-local modification to the standard Klein-Gordon theory, we analyzed the interaction between stationary point-like scalar charges for three representative cases: $M = 0$, $M = 2m$, and $M > 2m$. For these same regimes, we also computed the modified scalar propagator in the presence of a single Dirichlet plane.

Furthermore, we studied the interaction energy and the corresponding force between the Dirichlet plane and a point-like scalar charge. For the cases $M = 0$ and $M = 2m$, we derived exact analytical expressions. In the case where $M > 2m$, the analysis was carried out numerically. In all configurations, we found that the image method fails to reproduce the correct interaction, confirming its inapplicability within the non-local model described by Eq.~(\ref{model1}).

Remarkably, we find that, in the massless limit of the Klein--Gordon field, the interaction force between a Dirichlet plane and a scalar charge coincides (up to a sign) with the force previously obtained in the framework of nonlocal electrodynamics for a perfectly conducting plane interacting with a point-like electric charge. This surprising correspondence points to a deeper structural similarity between nonlocal scalar and gauge field theories in the presence of boundary conditions. It is important to stress that this agreement is far from trivial and does not persist even in simpler higher-derivative theories, such as Lee--Wick-type models \cite{BC15,BC16}.

As a direction for future work, it would be interesting to extend this analysis to the case of a semi-transparent Dirichlet plane~\cite{GTFABFEB}, using the same non-local scalar field model~(\ref{model1}). This may provide further insights into how boundary transparency affects non-local interactions.

\begin{acknowledgments}
L.H.C.B. acknowledges the financial support from Conselho Nacional de Desenvolvimento Científico e Tecnológico (CNPq) and Fundação de Amparo à Pesquisa do Estado de Minas Gerais-FAPEMIG (Project No. APQ-06536-24). C. Filgueiras acknowledges the financial support from FAPEMIG (Project No.  APQ 02226/22) and CNPq (Project No. 302783/2025-3). F. A. Barone acknowledges support from CNPq under Grant No. 313426/2021. A. A. Nogueira acknowledges full support from PROEPD/UDESC.
\end{acknowledgments}



\end{document}